\def\shownotes{1}
\def\conference{0}
\def\addproofs{1}

\ifnum\conference=1
\documentclass[letterpaper,twocolumn,10pt]{article}

\usepackage{usenix-2020-09}
\usepackage{hyperref}
\usepackage{cite}
\usepackage{amsthm}
\usepackage{amsmath}
\usepackage{txfonts}
\usepackage{wasysym}
\usepackage{mathtools}
\usepackage{xcolor}
\else
\documentclass{article} 
\usepackage{fullpage}
\usepackage{authblk}
\usepackage[draft]{hyperref}
\usepackage{amsthm}
\usepackage{amsmath}
\usepackage{txfonts}
\usepackage{wasysym}
\usepackage{mathtools}
\usepackage{amssymb}
\usepackage[table]{xcolor}
\fi

\usepackage{cite}
\usepackage{colortbl}

\usepackage{cite}
\usepackage{csvsimple}

\usepackage{graphicx}
\usepackage{microtype}

\usepackage{framed}
\usepackage{multirow}
\usepackage{color}

\usepackage{csquotes}
\usepackage{marginnote}

\usepackage[inline]{enumitem}
\setlist[enumerate,1]{label={(\arabic*)}}
\setlist{noitemsep}

\newtheorem{theorem}{Theorem}
\newtheorem{definition}{Definition}
\newtheorem{lemma}{Lemma}

\ifnum\shownotes=1
\newcommand\com[2]{{\marginpar{\bf\color{blue}{#1 : }{\color{blue}{#2}}}}}
\newcommand\todo[1]{{\bf\color{red}{TODO : }}{\color{red}{#1}}}
\else
\newcommand\com[2]{}
\newcommand\todo[1]{}
\fi

\ifnum\conference=1
\newcommand{\myparab}{\textbf}
\else
\newcommand{\myparab}{\paragraph}
\fi

\newcommand{\id}{\operatorname{id}}

\DeclareMathOperator{\Exp}{\mathbb{E}}

\newcommand{\cvr}{\mathsf{cvr}}

\newcommand{\disqual}{\mathtt{Disqual}}
\newcommand{\disagree}{\mathtt{Disagree}}
\newcommand{\omission}{\mathtt{Omission}}
\newcommand{\conflict}{\mathtt{Conflict}}
\newcommand{\act}{\mathrm{act}}
\newcommand{\obs}{\mathrm{obs}}

\newcommand{\size}{\mathsf{S}}
\newcommand{\winner}{\mathsf{W}}
\newcommand{\loser}{\mathsf{L}}
\newcommand{\cW}{\winner}
\newcommand{\cA}{\mathsf{A}}
\newcommand{\cB}{\mathsf{B}}
\newcommand{\candidates}{\mathcal{C}}

\newcommand{\disc}{\mathsf{D}}

\newcommand{\iter}{\mathsf{iter}}
\newcommand{\ballot}{\mathbf{b}}

\newcommand{\consistent}{\textsc{Consistent}}
\newcommand{\inconclusive}{\textsc{Inconclusive}}
\newcommand{\RLA}{\mathsf{RLA}}
\newcommand{\CompA}{\mathsf{CompeteA}}
\newcommand{\Stop}{\mathsf{Stop}}
\newcommand{\FirstStop}{\tau_{\Stop}}

\newcommand{\Riskdelta}{\mathsf{Risk}_\delta}
\newcommand{\risk}{\alpha}
\newcommand{\Criterion}{\mathsf{R}}

\newcommand{\Auditor}{\mathtt{MAudit}}
\newcommand{\CAuditor}{\mathtt{Judge}}

\newcommand{\Environment}{\mathtt{Env}}
\newcommand{\Stopmu}{\mathsf{Stop}_{\mu}}
\newcommand{\Stopdelta}{\mathsf{Stop}_{\delta}}
\newcommand{\Criterionmu}{\mathsf{R}_{\mu}}
\newcommand{\Criteriondelta}{\mathsf{R}_{\delta}}
\newcommand{\Identifiers}{\mathtt{Idents}}
\newcommand{\Claim}{\texttt{Claim}}
\newcommand{\Binomial}{\textsl{Bin}}
\newcommand{\truth}{\mathcal{T}}
\newcommand{\pred}{\mathcal{P}}
\newcommand{\struth}{T}
\newcommand{\spred}{P}
\newcommand{\rv}[1]{{\mathrm{#1}}}

\newcommand{\interpretations}{\ensuremath{\mathcal{I}}}
\newcommand{\tinterpretation}{\mathsf{T}}
\newcommand{\cinterpretation}{\mathsf{P}}
\newcommand{\R}{\mathbb{R}}
\newcommand{\N}{\mathbb{N}}
\newcommand{\vceil}{{\scriptsize{\CIRCLE}}}
\newcommand{\vfloor}{{\scriptsize{\Circle}}}
\newcommand{\vaverage}{{\scriptsize{\LEFTcircle}}}
\newcommand{\sure}{+}

\begin{document}

\title{The Decisive Power of Indecision:\\ Low-Variance Risk-Limiting
  Audits and Election Contestation\\ via Marginal Mark Recording\ifnum\conference=0
  \footnote{This is a revised and extended version of an article of the same name at USENIX Security 2024.}
  \fi
  }
\author{Benjamin Fuller, Rashmi Pai, Alexander Russell\\University of Connecticut -- Voting Technology Center\\{\tt \{benjamin.fuller,rashmi.pai,alexander.russell\}@uconn.edu}}

\ifnum\conference=1
\maketitle

\abstract{
Risk-limiting audits (RLAs) are techniques for verifying the outcomes
of large elections. While they provide rigorous guarantees of
correctness, widespread adoption has been impeded by both efficiency
concerns and the fact they offer statistical, rather than absolute,
conclusions. We attend to both of these difficulties, defining new
families of audits that improve efficiency and offer qualitative
advances in statistical power.

Our new audits are enabled by revisiting the standard notion of a
cast-vote record so that it can declare multiple possible mark
interpretations rather than a single decision; this can reflect the
presence of marginal marks, which appear regularly on hand-marked
ballots. We show that this simple expedient can offer significant
efficiency improvements with only minor changes to existing auditing
infrastructure. We consider two ways of representing these marks, both yield risk-limiting comparison audits in the formal sense of Fuller, Harrison, and
Russell (IEEE Security \& Privacy 2023).

We then define a new type of post-election audit we call a
\emph{contested audit}.  These permit each candidate to provide a
cast-vote record table advancing their own claim to victory. We prove
that these audits offer remarkable sample efficiency, yielding control
of risk with a constant number of samples (that is independent of
margin). This is a first for an audit with provable soundness. These
results are formulated in a game-based security model that specify
quantitative soundness and completeness guarantees. These audits provide a means to handle
contestation of election results affirmed by conventional RLAs. 
}

\else
\author{Benjamin Fuller, Rashmi Pai, Alexander Russell\\University of Connecticut -- Voting Technology Center\\{\tt \{benjamin.fuller,rashmi.pai,alexander.russell\}@uconn.edu}}

\maketitle
Risk-limiting audits (RLAs) are techniques for verifying the outcomes
of large elections. While they provide rigorous guarantees of
correctness, widespread adoption has been impeded by both efficiency
concerns and the fact they offer statistical, rather than absolute,
conclusions. We attend to both of these difficulties, defining new
families of audits that improve efficiency and offer qualitative
advances in statistical power.

Our new audits are enabled by revisiting the standard notion of a
cast-vote record so that it can declare multiple possible mark
interpretations rather than a single decision; this can reflect the
presence of ambiguous marks, which appear regularly on hand-marked
ballots. We show that this simple expedient can offer significant
efficiency improvements with only minor changes to existing auditing
infrastructure. We establish that these ``Bayesian'' comparison audits
are indeed risk-limiting in the formal sense of (Fuller, Harrison, and
Russell, 2022).

We then define a new type of post-election audit we call a
\emph{contested audit}.  These permit each candidate to provide a
cast-vote record table advancing their own claim to victory. We prove
that these audits offer remarkable sample efficiency, yielding control
of risk with a constant number of samples (that is independent of
margin). This is a first for an audit with provable soundness. These
results are formulated in a game-based security model that specify
quantitative soundness and completeness guarantees. Finally, we
observe that these audits provide a direct means to handle
contestation of election results affirmed by conventional RLAs.  \fi

\section{Introduction}

\emph{Risk-limiting audits} (RLAs) are methods for verifying the
outcome of a large-scale election. Developed by the academic and
election community over that last two
decades~\cite{lindeman2012gentle},
RLAs offer significant efficiency improvements over the burden of a
full hand recount while providing rigorous guarantees of correctness.
They can also support a variety of social choice
functions 
and election organizations (see Section~\ref{sec:rel work}). 
The framework posits reasonable physical assumptions under which elections---involving diverse interactions among multiple
untrusted human and electronic processes---can be authoritatively
protected from errors or deliberate falsification in tabulation, vote
aggregation, and reporting. Finally, the framework supports
\emph{public verification}: third-party observers can certify the
outcome of the audit~\cite{national2018securing}.

RLAs are only defined in settings with voter-verified ground truth,
typically furnished by hand-marked paper
ballots.\footnote{\href{https://verifiedvoting.org/verifier}{Verified
    Voting} summarizes the adoption of such methods in the U.S.}  Such
marked ballots determine the \emph{ground-truth outcome of the
  election}; of course, any particular tabulation of the
ballots---generally carried out by electronic tabulators---determines
a potentially different outcome~\cite{lindeman2012gentle}. RLAs are designed to detect
disagreement between the ground-truth outcome and a given tabulated
outcome except with a prescribed, concrete failure probability called
the ``risk'' of the audit. The word ``outcome'' here refers to the
winner(s) of the election rather than the exact vote
totals. 

  RLA adoption in large democracies has been uneven. In the United
  States, RLAs have been advocated by area
  experts~\cite{national2018securing}, the
  \href{https://www.intelligence.senate.gov/publications/russia-inquiry}{US
    Senate}, and the 2014
  \href{https://bipartisanpolicy.org/the-presidential-commission-on-election-administration/}{Presidential
    Commission on Election Administration}.  However, only a handful
  of states run such audits, with most adoption occurring in
  the last few years.  A major obstacle to widespread adoption is
  efficiency:
  RLAs require manual interpretation of a sampled set of ballots whose
  size grows as a function of the margin of the race to be
  audited. Thus, circumstances with small margins may require
  examination of many thousands of ballots and, unfortunately,
  election logistics and planning must prepare for such onerous
  outcomes. (As an example, a conventional ballot polling audit
  typically requires over 10,000 ballot samples for a 2\% margin and $50,000$ samples at $1\%$ margin~\cite{zagorski2021minerva}.)  A
  related challenge is that high-efficiency RLAs
  necessarily provide (only) statistical guarantees; common risk
  values in practice are on the order of $5\%$ (\href{https://www.coloradosos.gov/pubs/elections/RLA/2024/presidential/riskLimit.html}{3\% in Colorado}, \href{https://sos.ga.gov/news/georgias-2022-statewide-risk-limiting-audit-confirms-results}{5\% in Georgia}, \href{https://elections.ri.gov/sites/g/files/xkgbur756/files/publications/Election_Publications/RLA/Rhode-Island-presidential-RLA-brief-report.pdf}{9\% in Rhode Island}, \href{https://www.elections.virginia.gov/resultsreports/election-security/rla/}{10\% in Virginia}). This invites an immediate
  criticism: a losing candidate whose loss has been reaffirmed by an
  RLA may argue that a 1-in-20 chance of error is unsatisfactory.  We
  engage with both of these concerns by defining two new families of
  post-election audits.

  \subsection{Our Results in Brief}
Our new approaches are enabled by revisiting the
  conventional understanding of a cast-vote record (CVR),  a
  declaration of how a ballot was counted in an election. CVRs are
  critical elements of standard ``comparison'' audits, which are the
  starting points for our new audits. We redefine CVR semantics to
  include \emph{explicit indication of marginal
    marks}~\cite{bajcsy2015systematic} by either asserting a
  collection of possible interpretations for a marginal mark or a
  predicted probability distribution of interpretations. Marginal
    marks are those for which voter intent is unclear. Aside from
    partially filled voting targets, this can include ``small lines, atypical
    ink or marker, and marks outside but near the voting
    target''~\cite{bajcsy2015systematic}. We leverage this
  declaration of indecision to define two new families of audits.

\myparab{Ballot-comparison risk-limiting audits with
  reduced variance and improved efficiency.} Recording a
\emph{predicted probability distribution} for the interpretation of
marginal marks in a CVR and reflecting this appropriately in the audit
leads to a new family of \emph{Bayesian risk-limiting audits}, which
can offer significant efficiency advantages and improved confidence in
the audit. In particular, in settings of practical interest this
reduces the standard deviation of completion times for standard
sequential statistical testing methods and hence reduces the number of
ballots that need to be drawn in the first (and typically only) phase
of standard multi-phase audits. For margins of $1\%$, this reduces the
number of ballots sampled by over 10\%; see Table~\ref{tab:km sizes}.
We also discuss such audits in a less expressive ``conservative''
setting where the CVR simply lists a set of possible interpretations. (One could consider interpretations with some minimum probability as shown in Figure~\ref{fig:cvr visual}.) 
These results are articulated in an adaptation of the formal
game-based model of Fuller, Harrison, and Russell~\cite{harrison2022adaptive}.

Aside from these efficiency improvements, such CVRs provide an
improvement in voter confidence and interpretability by explaining
discrepancies. Specifically, conventional CVRs will frequently
disagree with marginal marks as, of course, there is no authoritative
conclusion for such a mark; this can cast doubt on the CVR and the
audit. A CVR permitted to assert the ambiguity of a mark provides an
immediate means for separating the common circumstances where the mark
is ambiguous from more disturbing discrepancies arising from real
failures of the audit (see further discussion in Section~\ref{sec:practical deployment}).
We remark that  \href{https://www.clearballot.com/}{Clearballot} publishes mark images with predicted probability distributions and allows citizens 
   to search for marginal marks; thus our method makes use of data that is already public in
   many jurisdictions (Clearballot indicates that $13$ US states use their technology).

\myparab{\emph{Competitive audits} involving multiple CVRs submitted
  by multiple interested parties, yielding a natural approach to
  \emph{election contestation}.} We define a novel notion of a
``competitive audit,'' in which multiple parties submit CVRs to
substantiate competing claims of victory. We develop a formal
cryptographic game-style framework for defining and analyzing such
audits. This modeling permits a strong adversary that may prepare all but one CVR,
choose which interpretation to return for a marginal mark, and whether
to suppress a sampled ballot.
We then show that this model leads to an \emph{extraordinarily efficient
  audit calling for evaluation of a fixed, constant number of
  ballots}, independent of margin.  Such strong efficiency guarantees
are impossible for conventional RLAs.  This new auditing framework
provides an approach to the problem of election contestation, where
the results of an RLA are challenged by a third party.
We discuss practical considerations for deploying these audits in Section~\ref{sec:practical deployment}.

\begin{figure*}[t]
  \centering
  \small
  \begin{tabular}{|c|c|c|} 
    \multicolumn{3}{c}{\large{Baseline CVR}}\\
    \multicolumn{3}{c}{}\\ \hline
    ID & Bugs & Daffy\\ \hline\hline
    1\cellcolor{yellow} & \cellcolor{black} & \\\hline
    2\cellcolor{yellow} & \cellcolor{black} & \\\hline
    3\cellcolor{yellow} & \cellcolor{black} & \\\hline
    4\cellcolor{yellow} &  & \cellcolor{black} \\\hline
    5\cellcolor{yellow} &  & \cellcolor{black} \\\hline
    6\cellcolor{yellow} &  & \\\hline
    7\cellcolor{yellow} & \cellcolor{black} & \\\hline
    8\cellcolor{yellow} & & \cellcolor{black}  \\\hline
    9\cellcolor{yellow} & & \cellcolor{black}  \\\hline
    10\cellcolor{yellow} &  & \\\hline
  \end{tabular}\qquad
  \begin{tabular}{|c|c|c|c|}
    \multicolumn{4}{c}{\large{Conservative CVR}}\\
    \multicolumn{4}{c}{}\\ \hline
    ID & Bugs & Daffy & No Vote \\ \hline\hline
    1\cellcolor{yellow} & \cellcolor{black} &  & \\\hline
    2\cellcolor{yellow} & \cellcolor{black!50} &  & \cellcolor{black!50} \\\hline
    3\cellcolor{yellow} & \cellcolor{black} & & \\\hline
    4\cellcolor{yellow} &  & \cellcolor{black} & \\\hline
    5\cellcolor{yellow} &  & \cellcolor{black!50} &  \cellcolor{black!50} \\\hline
    6\cellcolor{yellow} &  \cellcolor{black!50} &  \cellcolor{black!50} & \cellcolor{black!50}\\\hline
    7\cellcolor{yellow} & \cellcolor{black!50} & \cellcolor{black!50} & \cellcolor{black!50} \\\hline
    8\cellcolor{yellow} & & \cellcolor{black}  & \\\hline
    9\cellcolor{yellow} & & \cellcolor{black}  & \\\hline
    10\cellcolor{yellow} &  & & \cellcolor{black} \\\hline
  \end{tabular}\qquad
  \begin{tabular}{|c|c|c|c|}
    \multicolumn{4}{c}{\large{Bayesian CVR}}\\
    \multicolumn{4}{c}{}\\ \hline
    ID & Bugs & Daffy & No Vote \\ \hline\hline
    1\cellcolor{yellow} & 1  &  & \\\hline
    2\cellcolor{yellow} & .72  & .02 & .26 \\\hline
    3\cellcolor{yellow} & .99 &     & .01 \\\hline
    4\cellcolor{yellow} &     & 1   &     \\\hline
    5\cellcolor{yellow} &     & .75   & .25    \\\hline
    6\cellcolor{yellow} & .12  & .38  & .5    \\\hline
    7\cellcolor{yellow} & .46  & .1 & .44 \\\hline
    8\cellcolor{yellow} &     & 1   &     \\\hline
    9\cellcolor{yellow} &     & 1   &     \\\hline
    10\cellcolor{yellow}&     & .02 & .98 \\\hline
  \end{tabular} 
  \caption{Generalization of conventional CVRs to the conservative
    setting, where multiple interpretations are declared, and the
    Bayesian setting, where one associates probabilities with each
    interpretation. Blanks cells in the Bayesian indicate probability
    $0$; for brevity, the ``No Vote'' columns reflect both the no-mark
    case and the overvote case (with marks for both candidates).}
\label{fig:cvr visual}
\end{figure*}

\subsection{A Detailed Survey of the General Framework, Related Work,
  and Our Results}

An RLA is carried out in the context of an election with
voter-verified ground truth. For concreteness, our discussion assumes
hand-marked paper ballots, which is the most common instantiation in
practice. The audit is intended to detect disagreement between the
ground-truth outcome of the election, as determined by the ballots
themselves, and the \emph{declared} outcome of the election resulting
from a tabulation of the ballots; we use the word ``declared''
throughout to emphasize that the tabulated outcome may be inconsistent
with the ground-truth outcome. Such an audit examines a collection of
physical ballots, usually drawn at random according to a convention
depending on the audit, and either concludes with $\consistent$,
indicating that the sampled ballots appear consistent with the
declared outcome, or $\inconclusive$, indicating that the audit has
not amassed sufficient evidence of consistency. An audit is said to
have \emph{risk} $\alpha$ (a number in the range $[0,1]$) if the
probability that it outputs $\consistent$ is no more than $\alpha$
when the declared outcome contradicts the ground-truth outcome.
\textsc{Inconclusive} outcomes usually call for a full hand
  recount; see the next subsection.

A trivial audit that always outputs $\inconclusive$ has zero risk, but
is obviously not useful in practice. This points to the importance of
\emph{completeness}---the additional guarantee that the audit outputs
$\consistent$ with high probability under favorable circumstances; we
discuss this in detail below. A straightforward hand recount of an
election
is itself a RLA audit with zero risk (modulo any errors in
manual interpretation).
More sophisticated RLAs randomly
draw and inspect a collection of ballots to make statistical
conclusions about the outcome. A general feature of such statistical
audits is that the number of samples scales with the margin of
victory; this reflects the natural intuition that a landslide should
be easier to statistically verify than a victory by slim margin. In
the most familiar setting of a simple, two-candidate,
first-past-the-post election, the empirical mean of the number of
votes for a particular candidate among a collection of
$\log(2/\alpha)/(2\epsilon^2)$ uniformly sampled ballots is correct to
within $\epsilon$ with probability $1-\alpha$. This simple
procedure---called ``ballot polling'' in the literature---yields a
risk-limiting audit that inspects
\begin{align}
  O\left(\frac{\log(1/\alpha)}{\mu^2}\right) \label{eq:polling}
\end{align}
ballots in expectation, where $\mu$ is the actual margin~\cite{lindeman2012gentle}.

The most efficient known approach for RLAs, known as a \emph{ballot
  comparison
  audit}~\cite{lindeman2012gentle,stark2009auditing,stark2010super,higgins2011sharper,ottoboni2018risk,stark2020sets,waudby2021rilacs,blom2021assertion},
requires a detailed tabulation report called a cast-vote record table
(CVR). A CVR declares a unique identifier and interpretation (that is,
a determination of the cast votes) for each ballot in the election.
This ballot-by-ballot record of the election naturally determines a
tabulation and outcome.  It also enables high-efficiency audits that
proceed by iterating the following experiment:
\begin{enumerate*}
\item The auditor picks a random identifier $\iota$ from the CVR table; 
\item the auditor finds a physical ballot with identifier $\iota$;
\item the auditor compares the interpretation of the physical ballot
  against the associated CVR entry.
\end{enumerate*}
Intuitively, iterations where the physical ballot matches the CVR
provide support for this declared record of the election, while
disagreements erode that support. Continuing the discussion in the
same two-candidate first-past-the-post setting, observe that if the
CVR table declares an incorrect outcome for the election, then at
least a $\mu/2$ fraction of entries of the CVR table must disagree
with the corresponding physical ballot; as above, $\mu$ is the scaled
margin of ground-truth victory. (The factor of two arises because a
single disagreement can both negate a vote for the declared winner and
supply a vote for the declared loser.) It follows that after
$2/\mu$ samples we expect to observe a discrepancy if the
outcomes differ, suggesting that this approach should yield a
risk-limiting audit with sample size
\begin{equation}
  \label{eq:comparison}
  O\left(\frac{\log(1/\alpha)}{\mu}\right)\,.
\end{equation}
(The $\log(1/\alpha)$ term increases the number of samples so that the
probability of observing an inconsistency is driven to at least
$1-\alpha$.) It is instructive to compare this
against~\eqref{eq:polling} above. Due to  the improved dependence on
$\mu$, in circumstances with small
margins ballot comparison beats ballot
polling.

\myparab{The completeness challenge.} The efficiency
landscape is more complicated than the discussion above (and the
asymptotic expression~\eqref{eq:comparison}) indicates. The
complication is the requirement of completeness---that is, that the
audit conclude with $\consistent$ when considering an accurate CVR
under favorable conditions. In practice, when a post-election RLA
concludes with $\inconclusive$, policy calls for either a complete
hand count or an additional stage of auditing (with a smaller risk
parameter to avoid significant amplification of risk from the
composite audit). See for example, \href{http://webserver.rilin.state.ri.us/Statutes/TITLE17/17-19/17-19-37.4.HTM}{Rhode Island Code 17-19-37-4} and \href{https://www.sos.state.co.us/pubs/elections/Resources/files/ElectionFactSheet-PostElectionAudits.pdf}{Colorado's fact sheet}. 
Thus a full accounting of efficiency must weigh the
costs and likelihood of $\inconclusive$ conclusions in the best-effort
settings that arise in practice.

To be more concrete, a ballot-comparison auditor simply wishing to
meet a prescribed risk limit $\alpha$ could adopt the framework above
with the convention that \emph{any} observed disagreement between the
CVR and a physical ballot causes the auditor to immediately terminate
the audit and output $\inconclusive$. If examination of
$\log(1/\alpha)/(2\mu)$ ballots indeed exposes no inconsistencies, the
audit can safely return $\consistent$ with only an $\alpha$
probability of error, as desired. For margin $\mu=.01$ and risk
$\alpha=.05$, for example, this requires inspection of $499$ ballots.
Now, if this ``strict'' auditor has the luxury of unambiguous ground
truth, a perfectly correct CVR, and consistently accurate ballot
interpretations, then it will indeed conclude the audit with
$\consistent$. However, discrepancies between the CVR and physical
ballots arise frequently in practice. (For example, the comparison RLA
run by the US state of Colorado for the 2020 presidential election
generated hundreds of discrepancies~\cite{CO-DiscReport}.)  These
discrepancies can result
from truly ambiguous voter intent, disagreement among the
interpretations of an auditing board, or operational errors such as
retrieval of an incorrect ballot. Estimates from prior work (e.g. \href{https://www.stat.berkeley.edu/~stark/Vote/auditTools.htm}{Stark's audit tools}; \cite{VoTeR2022}) suggest
discrepancy rates as large as $0.5\%$, though the actual
rate of incidence depends on various features of the
election.

These considerations have led to the development and adoption of
audits (and associated statistical tests) that can tolerate such
anticipated errors. For example, adopting the same parameters ($5\%$
risk, $1\%$ margin) but with the introduction of a $.01\%$ rate of
marks being inadvertently dropped during interpretation, the standard
Kaplan-Markov test calls for a sample size of $1220$ ballots in order
to terminate with probability 95\%.\footnote{This estimate uses
  McBurnett's tool
  \href{https://bcn.boulder.co.us/~neal/electionaudits/rlacalc.html}{rlacalc}
  with standard rates of errors in the literature drawn from Lindeman
  and Stark~\cite{lindeman2012gentle}.  These are $o_1, u_1=.001$ and
  $o_2, u_2=.0001$, see discussion in Section~\ref{sec:efficiency}.}
This is over twice the sample size arising with no errors ($499$,
  as indicated above) and illustrates that even small
error rates have a significant impact on required sample size; we remark,
additionally, that the incidence of sporadic errors in an otherwise
correct CVR or errors in ballot interpretation are the cause of
variance in the stopping time for such sequential statistical
tests. As audits are typically conducted in phases---drawing a
collection of ballots to be inspected at once and calibrated so that
each phase is likely to be the last---variance also directly
influences efficiency as it forces large phases.

These concerns motivate our exploration of more expressive CVRs. In
particular, we note that the party (or tabulator) preparing the CVR
can identify marginal marks during tabulation that are likely to lead
to errors during a comparison audit. Reflecting these marginal marks
on the CVR offers a new dimension of optimization for the auditing
rule.
As a matter of bookkeeping, we will consider CVRs that may list
multiple interpretations for each ballot or a full predicted
probability distribution of interpretations. (A traditional CVR
corresponds to the case when the size of each set of interpretations
is $1$.) We apply this to develop both conventional audits with
improved efficiency and define a new class of competitive audits,
providing both high efficiency and a mechanism for contesting
conventional audits.

Enriching CVRs  to reflect marginal marks can also improve confidence in the
audit, as it can distinguish avoidable discrepancies arising from
ambiguity from unexplained auditing failures; see Section~\ref{sec:practical deployment}.

\subsubsection{Improved Efficiency Risk-Limiting Audits via Bayesian
  CVRs} 
  We show how to adapt standard ballot comparison audits to the
setting of \emph{Bayesian CVRs} that declare a predicted distribution
of ballot interpretations.
We indicate how to formulate the standard notions of discrepancy and
margin in this setting and observe that this simple change has
significant efficiency ramifications.  Using discrepancy and
marginal rates from practice we show that such audits reduce
sample size standard deviation by $14\%$ compared to a traditional audit at a margin of
$1\%$ (see Table~\ref{tab:km sizes}). We also develop a variant of
this audit that uses a CVR that merely reflects the \emph{possible}
interpretations of each ballot (rather than a full probability
distribution). We call this a \emph{conservative CVR}. While this does not provide the full benefits of the
Bayesian approach, it may be simpler to deploy in practice since no
probability estimation is required. It also considers a stronger adversarial model where the ballot interpretation of the auditor is adversarially controlled.

We adopt to our setting the cryptographic game-style modeling for RLAs
developed in~\cite{harrison2022adaptive}, and prove that the
new audit is risk-limiting. We then carry out simulations to
show that the most popular sequential statistical test used for RLAs
(the Kaplan-Markov test) can take advantage of the resulting notion of
discrepancy to provide more efficient RLAs.

\subsubsection{Competitive Audits and Election Contestation}
Our second contribution is a new class of post-election audits that we
call \emph{competitive RLAs}.  In a competitive RLA each candidate has
an ``advocate'' that is allowed to examine ballots (under supervision
to ensure ballots are not destroyed or modified).  For each candidate
$\cA_1,\ldots, \cA_k$, their advocate produces a CVR that may declare
different ballot interpretations and winners; the intent is that the
advocate for candidate $\cA$ files a CVR that is favorable to
$\cA$. In particular, if $\cA$ won the election, the CVR (filed by the
advocate for $\cA$) should establish that; in a weaker sense, even if
$\cA$ did not win, the advocate may wish to file a CVR that prevents
other candidates that did not win the election from convincingly
claiming that they did. For simplicity, we formulate this audit in the
``conservative'' setting, where CVRs may list multiple interpretations
for each ballot but do not record a probability distribution on these
various possible interpretations. (It is certainly possible to
formulate this in the setting with a Bayesian CVRs, see end of
Section~\ref{sec:contested audits}.) For elections with very small margins, an interpretation of marginal votes may be necessary to determine which 
candidate receives the most votes.  We are only able to prove results about candidates
 whose winning and losing status
is true regardless of interpretation of marginal marks. Definitionally, an election thus at most one winner, and some candidates
may neither win nor lose.

Assuming that at least one submitted CVR only lists interpretations
that are consistent with the ballots and the election has a winner,
the auditor will identify the correct winner using a constant number
of ballots. Furthermore, if some submitted CVR lists the exact list of
possible interpretations for each ballot, no losing candidate can be
identified as the winner.
We emphasize that the number of ballot samples required by this
approach is a constant that not scale with the margin. (The constant
itself is a function of how robust the audit should be in the face of
errors by the advocates.)  

The basic idea of the audit is straightforward: Faced with a pair of
contradictory CVRs---$\cvr_1$ and $\cvr_2$---there must be at least one
ballot on which these CVRs completely disagree: examining this single
ballot then provides an immediate strike against one of CVRs. Indeed,
a strict auditor can use a generalization of this approach to settle
the competing claims made by a collection of $k$ CVRs with just $k-1$
ballots. (That is, the auditor immediately removes from consideration any CVR found to be inconsistent with an inspected ballot.)
Such a stringent auditor will not be appropriate in practice, which
must account for a small number of errors on the part of the advocates
(and the $\cvr_i$) or the possibility that ballots are lost, say,
during the audit. Even accounting for such phenomena, our approach
yields audits with remarkable efficiency. (As mentioned earlier, the
complexity of the audit scales with the desired robustness in the face
of such advocacy errors.)

The most natural and practical instantiation of our techniques is
providing an efficient adjudication process for disagreements that
result after the parallel scanning of ballots, for example, using the
OpenScan system~\cite{wang2010openscan}. The idea of the OpenScan
system is for advocates to place cameras above a physical scanner and
create their own CVRs without having physical access to the ballots.
Their system does not have a natural adjudication mechanism. 
Competitive audits provide that adjudication with a constant query ballot complexity. 

\myparab{Election contestation.} Competitive RLAs provide an
immediate approach to election contestation. Specifically, consider a
situation where a candidate wishes to dispute the conclusions of
conventional RLA (that is, the candidate claims that an invalid
election was not detected by the RLA). In this case, the candidate (or
their advocate) can be permitted to examine the ballots and produce a
competing CVR for the election (that, presumably, will claim that the
candidate is the victor). The CVR originally produced for the election
(and reaffirmed by the original RLA) and the candidate's CVR can then
be treated by the competitive audit mechanism. This has a number of
notable features: (i.) for some integer $t$, the audit can rigorously settle the dispute
with risk $2^{-\theta(t)}$ after examining only $t$ ballots---thus a
small constant number of queries, independent of margin, are
sufficient; (ii.)  this provides an interesting avenue for election
policy: for example, the cost and effort of the audit could be borne
by the contesting party. See more discussion in Section~\ref{sec:practical deployment}.

\myparab{Adversarial modeling.}
We consider a cryptographic game-based
model for analysis. The model places the ballots (and, in conservative
and competitive audits, the choice of the returned interpretation) in
the hands of an adversarial environment, thus reflecting
various adverse circumstances that challenge practical audits. Our
  treatment universally quantifies over the ground-truth
  results (and, in particular, the marks appearing on the ballots
  themselves) and tabulated cast-vote records; in this sense they
  apply to all elections.
  We remark that in terms of the formal modeling any universally
  quantified object can be seen as adversarially controlled.

\ifnum\conference=0
 \subsubsection{Extensions and Future Work}
 \myparab{Ballot Polling with Marginal Marks}
  If one incorporated ballots with marginal marks into the polling process, this would require an ability to compute a questionable diluted margin without a CVR. Since polling is sound for any margin, this change retains soundness, it is unlikely to improve efficiency.

 \myparab{Batch Comparison with Marginal Marks}
 Batch comparison is an auditing process that involves randomly selecting batches of ballots to hand count (rather than individual ballots), and computing an observed discrepancy between the hand-counted ballots within the batch, and the tabulated totals of the ballots in the batch. For this process, we see no obstacles in incorporating ballots with marginal marks; our techniques directly translate into the batch setting. 
 \fi

\subsection{Other Related Work}
\label{sec:rel work}
Stark~\cite{starkconservative} defined RLAs in 2008. The
  framework was generalized and sharpened over the next
  decade~\cite{bernhard2021risk,verifiedvotingprinciples,Hall2009,
    lindeman2012gentle} leading to a variety of specific methods,
  often optimized certain practical settings~\cite{lindeman2012gentle,
    stark2010super, lindeman2012bravo, starkconservative, checkoway,
    starkcast, blom2022first}. There are four major directions in
  current research
\begin{enumerate*}\item supporting varied social choice functions~\cite{stark2020sets,blom2021assertion}, \item
  managing multiple
  races~\cite{stark2009auditing,stark2010super,higgins2011sharper,ottoboni2018risk,stark2009efficient}, \item
  sharpening risk
  estimates~\cite{higgins2011sharper,lindeman2012bravo,ottoboni2019bernoulli,waudby2021rilacs,Stark:Conservative,stark2023alpha,starkconservative,checkoway,banuelos2012limiting,
    zagorski2020athena} and \item implementation
  issues~\cite{verifiedvotingprinciples,Hall2009,bernhard2021risk,harrison2022adaptive,
    RLAWorkbook, rla-working-group, zagorski2021minerva,
    glazer2020bayesian, benaloh2021vault}.\end{enumerate*}

The most closely related work to our competitive setting is a recent,
independent article of Jones et al.~\cite{jones2022scan}. One can view
our competitive audits as an analog of a ``multi-prover interactive
proof system'' consisting of competing and potentially malicious
``provers'' who are attempting to convince a ``verifier'' of the truth
(or falsity) of a statement (reflecting the complexity-theoretic
setting of~\cite{reif1984complexity,Kiwi:2000aa}).  Jones et al.\ were
inspired by the ``cooperative'' version of this same framework in
which a collection of cooperating, but potentially malicious, provers
attempt to convince a verifier of the truth of a statement;
in fact, this cooperative version is far more common in the
complexity-theoretic literature (reflecting the standard setting
of~\cite{babai1991non}). Jones et al.\ develop an auditing approach
designed to catch inconsistencies across multiple scans used to create
(multiple) CVRs. In our auditing context, their techniques focus on
detecting circumstances where at least one of the scans (provers) is
dishonest.  Our techniques focus on deciding who to trust in a
situation where at least one scan is honest. The two approaches can
thus naturally compose.

\ifnum\conference=0
The key element in Jones et al.\ approach is a secret, random
permutation used to shuffle the ballot orderings between the distinct
scans. Discrepancy is calculated by comparing the outcome for selected
ballots between the scans.
The procedure provides similar efficiency guarantees: a constant
number of ballots, independent of margin. We remark that their
goal---detecting a failed scanner when the operators of the scanners
may be colluding---appears to require stronger assumptions than are
necessary for our competitive audits.  Most importantly, they require
that distinct ballots carrying the same votes are indistinguishable
and (as indicated above) the ability to physically instantiate a
private, random shuffle that will be reflected in scan order.
\fi

\myparab{Organization.}
We discuss practical deployment considerations in
  Section~\ref{sec:practical deployment}. We then cover notational
preliminaries and define CVRs and ballots with multiple
interpretations in Section~\ref{sec:preliminaries}, cover Bayesian
audits in Section~\ref{sec:questionable audits}, and Conservative
audits in Section~\ref{sec:conservative}. Finally, we evaluate
efficiency of the Bayesian and conservative methods in
Section~\ref{sec:efficiency} and describe competitive audits in
Section~\ref{sec:contested audits}. 
\ifnum\conference=1
Proofs are relegated to the online version~\cite{fuller2024decisive}.
\fi

\section{Practical Deployment Considerations}
\label{sec:practical deployment}
\myparab{Bayesian and conservative audits}
The deciding factor in adopting RLA modifications is how they
  impact the balance of trust in the election outcome and auditing
  effort. The essential trust assumptions of these new audits---that
  physical ballots faithfully represent voter intent and are available
  for interrogation---are identical to classical risk-limiting
  audits. However, we note that Bayesian and conservative audits have
  the advantage that marginal marks are directly reflected in the
  cast-vote record, a benefit for public perception and
  interpretability.  For years,
Colorado
    has published a discrepancy report~\cite{CO-DiscReport} explaining their guess of each
  nonzero discrepancy to improve public confidence in the audit. In particular, marginal marks which would have
  resulted in a consistency failure in a conventional audit are now
  explicitly indicated and reconciled with ballots; thus they serve to
  add to public confidence in the audit rather than detract from it.
  As for public interpretability, conservative audits appear superior to Bayesian audits 
  in the sense that the straightforward role played by a ``possible''
  interpretation in a conservative audit is clearly preferable from
  the standpoint of intelligibility to the fractional discrepancies
  arising from ``likelihood'' estimates. As conservative audits offer
  most of the statistical improvements of Bayesian audits in our
  numerical estimates, their relative simplicity may outweigh the
  small improvements in efficiency. We remark that these new audits
  can rely on the same statistical tests that underly conventional
  comparison audits.

  As for the challenge of transitioning procedures and equipment to
  support these audits, the substantive changes pertain to
  tabulation equipment (responsible for producing the CVRs) and RLA
  software. In particular, the task of the ``audit board'' (which
  evaluates the marks on sampled ballots) is identical to that of a
  conventional audit: they are still responsible for settling on a
  single interpretation for each sampled ballot. Both audits demand a
  richer convention for CVR, and hence changes to
  tabulator output and RLA administration software. As modern
  tabulators typically collect full ballots scans, the changes
  required to generate such CVRs would require only a firmware upgrade
  (rather than hardware changes). Again, the task here may be
  simplified by the conservative setting, which requires no explicit
  likelihood estimation.  Changes to RLA administration software are
  minimal, especially considering that existing statistical tests can 
  be applied. Our simulation~\cite{pai2024questionable} builds on Harrison's implementation with minimal changes~\cite{harrison2023adaptive}.
  We remark that the recent
  \href{https://www.eac.gov/sites/default/files/TestingCertification/Voluntary_Voting_System_Guidelines_Version_2_0.pdf}{VVSG
    2.0} federal voting system guidelines require ballot imprinting,
  even on voter-facing tabulators. This is an important enabling
  feature for ballot comparison audits, so adoption of these
  guidelines will significantly improve the general feasibility of
  comparison audits in the United States.

Compare the cost of
  conservative audits against the natural approach that calls for
  auditors to  adjudicate every marginal ballot as a part of the
  initial tabulation. As an illustrative example, Louisiana is the
  $25$th state by population with roughly $2$M votes for president in
  2020. Using the $.5\%$ marginal mark rate~\cite{VoTeR2022}, one
  would have to adjudicate $10$K ballots, in addition to what is required for the RLA. At $1\%$ margin
  ballot comparison requires examining $1$K ballots (Table~\ref{tab:km
    sizes}).

\myparab{The competitive setting.}
Competitive audits have roles and procedures that
have no analog in conventional RLAs.  Thus, they demand careful 
consideration before deployment. We outline two
different implementations and discuss the trust and efficiency
tradeoffs. The first implementation we discuss provides all of the
guarantees---including software independence---of a classical RLA
while providing an efficient contestation mechanism under two assumptions:
\begin{enumerate}
\item That there is an actual winner (regardless of how marginal marks are interpreted), and
\item An advocate for the winning party contests an incorrect outcome. 
\end{enumerate} The
contestation phase has a risk that reduces exponentially in the number of ballots pulled.
 We also discuss standalone instances of competitive audits,
including an existing framework that supports CVR creation by
candidates without directly handling ballots.

Consider the following RLA procedure with an additional contestation
phase:
\begin{enumerate}
\item The state's tabulators are run on the ballots, producing $\cvr_{\mathtt{state}}$ that declares candidate $\winner$ to be the winner. 
\item A conventional ballot comparison RLA is run on $\cvr_{\mathtt{state}}$ with risk $\alpha$.
\item If the RLA outputs $\inconclusive$, existing legal steps are followed such as a recount. 
\item Otherwise, candidates can contest the results of the
  audit. (Whether such contestations are permitted, or what evidence
  might be necessary to justify them, is a matter of election law.)
\item A candidate $\cA$ permitted to contest the election may inspect
  the physical ballots (under appropriate supervision to prevent
  tampering) to assemble $\cvr_\cA$, an alternate CVR. Various
  procedures could be used to create $\cvr_\cA$; for example, (i.) a
  complete hand count of all ballots, (ii.) use of alternative
  tabulators trusted by the candidate, (iii.) a hybrid approach, where
  the candidate uses $\cvr_{\mathtt{state}}$ with the exception of a
  few precincts of the election that they believe are suspicious; these
  are hand counted.
\item Finally, the competitive audit is run on all of the available CVRs (i.e., $\cvr_{\mathtt{state}}$ and $\cvr_{\cA}$ for any candidate $\cA$ that assembled a CVR).
\end{enumerate} This audit provides the guarantees of a
conventional RLA and provides a rigorous approach to contestation.
The audit  places the  vast majority  of the  work of  the competitive
audit  on the  party that  believes there  is an  error and  wishes to
correct it. The effective risk of  the combined audit can be driven to
a small  constant, e.g., one  part in 10,000,  with a small  number of
samples (assuming candidates can create accurate CVRs). We remark that
such an audit  continues to protect a third party  (with only a vested
interest in  the correct  outcome of the  election) from  an incorrect
result  with  risk  $\alpha$;  in particular,  when  the  state's  CVR
correctly asserts  the correct  winner, the contestation  phase cannot
change the outcome. On the other hand, an interested party (wishing to
ensure that a candidate of interest is not incorrectly declared a
loser of the election) can guarantee risk close to zero. It is a legal matter when to allow a contestation, our method provides a tool to 
decide when an advocate has gathered sufficient evidence.

One can also consider a standalone competitive RLA. The
  procedure follows the prescription above without the conventional
  RLA. (As with the audit above, supervision is required to protect
  ballot integrity.) If candidate $\cA$ is the winner and indeed
  submits a correct $\cvr_{\cA}$, $\cA$ will win the competitive audit
  regardless of the procedures or software adopted by other candidates
  (or state) to assemble their CVRs.
  We remark that such tabulations significantly change the guarantees
  offered to a third party with a vested interest only in the correct
  outcome of the election. With this audit, such a third party can
  only be guaranteed a correct outcome if the state's
  tabulation is correct or the true winning candidate participates in the competitive audit.

  Such competitive audits provide a rigorous approach to reconciling
  the results generated by the existing
  Openscan~\cite{wang2010openscan} system. This system permits each
  candidate to install their own camera over a scanning bed on which
  each ballot is rested for data collection. Adapting slightly to our
  framework, advocates do not touch ballots, but can be in the room to
  ensure each ballot is scanned exactly once. This has the advantage
  that all data is collected in one pass, and advocates use their own
  hardware and software to collect and process ballot images. Of
  course, the state itself may have an installed camera that produces
  an official outcome. The competitive audit now provides a rigorous
  guarantee to each candidate: if they won the election (and
  themselves generate a faithful CVR), they can be guaranteed victory
  independent of the software used by other candidates or the
  state. This assumes, of course, that each physical ballot is
  processed by their camera.

The above deployments require physical access to ballots---this
  is unavoidable for software-independent guarantees.  In either
  deployment, candidates shoulder auditing responsibilities (borne
  entirely by the official auditors in a traditional RLA). This is one
  advantage of the first deployment scenario: it limits candidate
  effort to only those circumstances where they suspect that a
  traditional RLA has failed.  However, competitive audits introduce a
  potential inequity:  Candidates with fewer financial resources may
  have less ability to muster a full retabulation for the competitive audit.

Our results assume a low rate of disagreement between the ``honest''
CVR and the set of interpretations that could be reasonably inferred by an audit board.
Preparing a CVR with a low error rate may require knowledge on mark adjudication and audit board behavior.

\section{Preliminaries; the Bayesian Setting}
\label{sec:preliminaries}

We use boldface to refer to ``physical'' objects, such as individual
ballots (typically denoted $\mathbf{b}$) or groups of ballots
(typically $\mathbf{B}$).
For a natural number $k$, we define $[k] = \{1, \ldots, k\}$ (and
$[0] = \emptyset$). We let $\Sigma = [-2, 2]$, a set of particular
significance as standard single-ballot discrepancy takes values in
this set; this is discussed in detail later in the paper.  For a set
$X$, let $X^*$ be the set of all finite-length sequences over $X$;
that is, $X^* = \{ (x_1, \ldots, x_k) \mid k \geq 0, x_i \in X \}$. In
particular, we let $\{0,1\}^*$ denote the collection of finite-length
bitstrings.
Finally, we define $X^\N$ to be the set
of all sequences $\{ (x_0, x_1, \ldots) \mid x_i \in X\}$.
We typically use script letters, e.g., $\mathcal{P}$, to denote
probability distributions. 
Making multiple interpretations explicit requires many new definitions.  A summary is in Table~\ref{tab:notation summary}.

\begin{table}[t!]
\centering
\ifnum\conference=1
\footnotesize
\begin{tabular}{l | r | p{2in}}
\else
\begin{tabular}{l | r | p{4.5in}}
\fi
Audit & Notation & Meaning\\\hline
\multirow{8}{*}{All}& $\mathbf{b}$ & Physical ballot\\
& $ \mathbf{B}$ & Ballot family\\
& $\mathbf{B}_\iota$ & Ballots with identifier $\iota$\\
& $\size$ & Number of ballots in election\\
& $\candidates$ & Candidates in the election\\
& $\alpha$ & Risk of audit\\
& $\gamma$ & Inflation factor of Kaplan-Markov when discrepancy encountered\\
\multirow{20}{*}{Bayes.}& $E$ & Election\\\hline\hline
& $\truth_{\mathbf{b}}$ & Probability distribution over auditor interpretation of ballot $\mathbb{b}$\\
& $\rv{I}_{\mathbf{b}}$ & R.V. with distribution $\truth_{\mathbf{b}}$\\
& $\tinterpretation_{\mathbf{b}}^{\vaverage}(\cA)$ & Expected number of votes for ballot $\mathbf{b}$ received by candidate $\cA$ if audited\\
& $E^\vaverage(\cA)$ & Expected number of votes across election received by candidate $\cA$ if hand counted\\
& $\pred_{r}$ & Predicted distribution of interpretations for ballot with identifier $r$\\
& $\cinterpretation^{\vaverage}_r(\cA)$ & Number of votes predicted for candidate $\cA$ on ballot with identifier $r$\\
& $ \mu_{\cA, \cB} $ & Bayesian diluted margin of $\cA$ with respect to candidate $\cB$\\
& $\mu_{\cA}$ & Minimum of Bayesian diluted margin across candidates\\
& $\cvr$ & Cast vote record containing identifiers and predicted interpretations for each ballot\\
& $\cvr(\cA)$ & Predicted number of votes for $\cA$ on $\cvr$\\
& $\mu_{\cvr}$ &  Bayesian margin declared by $\cvr$\\
& $\disc_{\cvr} $ & Discrepancy of $\cvr$ with respect to election $E$\\\hline\hline
\multirow{18}{*}{Cons.} & $\struth_{\mathbf{b}}$ & Set of possible interpretations on $\mathbf{b}$\\
& $\struth^{\vceil}_{\mathbf{b}}(\cA)$ &Maximum interpretation for $\cA$ on  $\mathbf{b}$\\
& $\struth^{\vfloor}_{\mathbf{b}}(\cA)$ &Minumum interpretation for $\cA$ on  $\mathbf{b}$\\
& $\struth^{\vceil}(\cA)$ &Most possible votes for $\cA$ in election\\
& $\struth^{\vfloor}(\cA)$ &Least possible votes for $\cA$ in election\\
& $ \mu_{\cA, \cB}^{\sure} $ & Conservative diluted margin of $\cA$ with respect to candidate $\cB$\\
& $\mu_{\cA}^{\sure}$ & Minimum of Conservative diluted margin across candidates\\
& $\mu_{\cvr}^{\sure}$ &  Conservative diluted margin in $\cvr^{\sure}$\\
& $\cvr^{\sure}$ & Conservative $\cvr$ declaring set of possible interpretations\\
& $    \cinterpretation^{\vfloor}_r(\cA)$ & Minimum predicted interpretation for $\cA$ \\
& $    \cinterpretation^{\vceil}_r(\cA)$ & Maximum predicted interpretation for $\cA$ \\
& $    \cvr^{\vfloor}(\cA)$ & Minimum possible votes for $\cA$ on $\cvr$\\
& $    \cvr^{\vceil}(\cA)$ & Maximum possible votes for $\cA$ on $\cvr$\\
& $\disc_{\cvr}^{\sure} $ & Conservative discrepancy of $\cvr$ with respect to election $E$\\\hline\hline
\end{tabular}

\caption{Summary of Notation}
\label{tab:notation summary}
\end{table}

\myparab{Formal definitions for ballots, elections, and cast-vote
  records.} For simplicity, we consider only audits of a single
first-past-the-post race with a set $\candidates$ of candidates. See
Stark~\cite{stark2010super,stark2020sets} for a detailed account of
how more complex elections can be reduced to this canonical
setting. We now set down the elementary definitions of elections,
manifests, and CVRs.  \textbf{The main definitional change is that a
  physical ballot induces a probability distribution of possible
  outcomes;} this models the important situation where a physical
ballot may have ambiguous markings that could reasonably be
interpreted in various ways by an auditing board or a tabulation
device.
  We explore two conventions for cast vote records to reflect this
  uncertainty: the first is the \emph{Bayesian} framework, which
  associates with each ballot a predicted probability distribution of
  possible readings (depending on the anticipated reading a particular
  audit board, say). We then formulate a simpler, \emph{conservative}
  model that posits a collection of interpretations for each ballot
  without assigning them probabilities. This reduces the complexity of
  producing cast-vote records, as one only needs to identify the
  \emph{possible} interpretations of a ballot rather than a
  comprehensive probability distribution.  In this conservative model,
  we consider quite strong adversarial behavior that assumes that the
  audit board always returns the ``worst-case'' interpretation allowed
  by the ballot.  (In fact, this is equivalent to replacing the
  probabilities listed on the CVR with a probability of $1$ for some
  interpretation that ``minimizes'' the margin for a declared winner.)
  To move as swiftly as possible to our results on conventional audits
  with marginal marks, we first lay out the definitions for the
  Bayesian case; we return to the definitions for the conservative
  case in Section~\ref{sec:conservative}.

\begin{definition}[Interpretations, ballots, and ballot families]
  \label{def:ballot-family} Let $\candidates$ be a set of candidates.
  An \emph{interpretation} is a function
  $I: \candidates \rightarrow \{0,1\}$; when $\candidates$ is
  understood from context, we write
  $\interpretations = \{ I: \candidates \rightarrow \{0,1\}\}$ for the
  collection of all interpretations. Such interpretations 
  to indicate the votes appearing for candidates on individual
  ballots.
  A \emph{ballot} is a physical object $\mathbf{b}$ with two
  properties:
  \begin{enumerate}
  \item The ballot $\mathbf{b}$ is \emph{labeled} with an indelible
    \emph{identifier} $\id_\mathbf{b} \in \{0,1\}^*$.
  \item The ballot $\mathbf{b}$ determines a ``ground truth''
    probability distribution $\truth_{\mathbf{b}}$ on the set of
    interpretations $\interpretations$. Thus
    $\truth_{\mathbf{b}}: \interpretations \rightarrow \R$ is a
    non-negative function on $\interpretations$ for which
    $\sum_{I \in \interpretations} \truth_{\mathbf{b}}(I) = 1$.
 \end{enumerate}
    We let
    $\tinterpretation_\mathbf{b}^{\vaverage}: \candidates \rightarrow
  \R$ denote the expected interpretation over the distribution
  $\truth_\mathbf{b}$: that is, for each $\cA \in \candidates$,
$
      \tinterpretation_{\mathbf{b}}^{\vaverage}(\cA) = \Exp[\rv{I}_\mathbf{b}(\cA)]\,,
  $
    where $\rv{I}_{\mathbf{b}}$ is a random variable distributed
    according to $\truth_{\mathbf{b}}$.

  A \emph{ballot family} $\mathbf{B}$ is a collection
  of ballots with the same candidates.  We let
  $\mathbf{B}_\iota = \{\mathbf{b} \in \mathbf{B} \mid
  \id_{\mathbf{b}} = \iota\}$ denote the subset of ballots with
  identifier $\iota$. When ballots identifiers are distinct across
  $\mathbf{B}$, so that $|\mathbf{B}_\iota| \leq 1$ for each $\iota$,
  we say that the family is \emph{uniquely labeled}.
\end{definition}

Ballots are typically intended to have an unambiguous
interpretation. This corresponds to the case where the distribution
$\truth_{\mathbf{b}}$ is supported on a single interpretation (taking
the value 1 at that interpretation and zero elsewhere).

\begin{definition}
  An \emph{election} $E$ is a tuple $(\candidates, \mathbf{B}, \size)$
  where $\candidates$ is a set of candidates, $\mathbf{B}$ is a ballot
  family, and $\size = |\mathbf{B}|$ is the total number of ballots.
\end{definition}


\begin{definition}[Election winners, losers, and margin]
  \label{def:election winner}
  Let $E = (\candidates, \mathbf{B}, \size)$ be an election. For
  a candidate $\cA \in \candidates$, define
  \begin{align*}
  E^\vaverage(\cA) =
  \sum_{\mathbf{b} \in \mathbf{B}}
  \tinterpretation_{\mathbf{b}}^\vaverage(A).
  \end{align*}
The \emph{margin of $\cA$ with respect to $\cB$} is defined to be
\[
  \mu_{\cA, \cB} = \frac{E^{\vaverage}(\cA) -  E^{\vaverage}(\cB)}{\size}.
\]
When we wish to emphasize that this notion of margin is defined with
respect to the expected values
$\tinterpretation_{\mathbf{b}}^\vaverage(\cdot)$ we refer to it as
\emph{Bayesian margin}. A candidate $\winner \in \candidates$ is the
\emph{winner} of the election $E$ if
$\forall \cA \in \candidates \setminus \{ \winner \}, \mu_{\cW, \cA}
>0$.  In this case, we define the \emph{margin}, denoted
$\mu$, of the election to be the minimum of these quantities:
\[
  \mu_E =  \min_{\cA \in \candidates\setminus \{\winner\}}\mu_{\cW, \cA}\,.
\]
Otherwise, the election does not determine a winner and we define
$\mu_E :=0$.
If a winner exists it is unique. A candidate $\loser$ is
called a \emph{loser} if there is a candidate $\cA$ for which
$E^{\vaverage}(\loser) < E^{\vaverage}(\cA)$.
\end{definition}

\myparab{Cast-vote record (CVR).}
A cast-vote record table (CVR) is an (untrusted) declaration of both
the ballots appearing in an election and the interpretations of the
ballots. 
\begin{definition}[Bayesian Cast-Vote Record Table]
  Let $\candidates$ be a set of candidates. A \emph{Bayesian Cast-Vote
    Record table (CVR)} is a sequence of pairs
$
    \cvr = \bigl((\iota_1,\pred_1), \ldots, (\iota_t,\pred_{\size_\cvr})\bigr)
  $
  where the $\iota_r$ are distinct indetifier bitstrings in $\{0,1\}^*$
  and each $\pred_r:\interpretations \rightarrow \R$ is a probability
  distribution on $\interpretations$ which we refer to as a
  ``prediction.'' Intuitively, such a $\cvr$ declares that, for each
  row $r$, the interpretation of the ballot labeled $\iota_r$ is given
  by the distribution $\pred_{r}$.
  
  For each candidate $\cA \in \candidates$ and
  $r \in \{1, \ldots, \size_\cvr\}$, we define
  $\cinterpretation^{\vaverage}_r(\cA) = \Exp[I(\cA)]$,
  the expected value of the vote for $\cA$ when $I$ is drawn according
  to the distribution $\pred_r$. We then define the total $\cvr$ declaration for candidate $\cA$ as
  \[
    \qquad \cvr(\cA) = \sum_{r=1}^{\size_\cvr} \cinterpretation^{\vaverage}_r(\cA)\,.
  \]

  \textbf{Declared winners and losers.} If there is a candidate
  $\cA \in \candidates$ with the property that $\cvr(\cA) > \cvr(\cB)$
  for every $\cB \in \candidates \setminus \{\cA\}$, then we say that
  $\cA$ is the \emph{declared winner} according to $\cvr$. A candidate $\cA$ for which there exists some $\cB$ such that $\cvr(\cA)< \cvr(\cB)$ is called a \emph{declared loser}.

  We use the following general language when referring to CVRs:
  \begin{enumerate}
  \item The bitstrings $\iota_r$ are \emph{identifiers} and we let
    $\Identifiers(\cvr)$ denote this set.
  \item The number $\size_\cvr$ is the \emph{size} of the CVR.
  \item The \emph{$r$th row}, denoted $\cvr(r)$, refers to the tuple
    $(\iota_r, \pred_r)$.
  \item Identifiers appearing in the CVR are distinct, so when
    convenient we reference $\pred_r$ and $\cinterpretation_r$ by
    identifier rather than row: that is, for an identifier $\iota = \iota_r$
    appearing in the CVR, we define $\pred_\iota := \pred_{r} $ and
    $\cinterpretation_\iota^{\vaverage} := \cinterpretation_r^{\vaverage}$.
  \end{enumerate}
\end{definition}

\begin{definition}[Bayesian declared margin]
  Let $\cvr$ be a Bayesian CVR with declared winner $\cW$ and size
  $\size_\cvr$.  The \emph{declared margin} of $\cvr$ is the quantity
    \[
      \mu_{\cvr} := \min_{\cA\in \candidates \setminus \cW}
      \frac{\cvr(\cW) - \cvr(\cA)}{\size_\cvr}\ge 0.
    \]
   
   \noindent
    If $\cvr$ has no declared winner, we define $\mu_\cvr := 0$. 
\end{definition}

\begin{definition}[Bayesian validity]
  Let $E=(\candidates, \mathbf{B}, \size)$ be an election and $\cvr$
  be a Bayesian CVR.  We say that $\cvr$ is \emph{valid} for $E$ if
  $E$ has a winner and $\cvr$ declares that winner.  We say that
  $\cvr$ is invalid for $E$ if it declares a winner $\cW$ that  does 
  not win $E$.
\end{definition} 

Discrepancy is a standard measure of how much a CVR overcounts
  $\winner$'s margin over other candidates $\cA\neq \winner$.
  Discrepancy naturally extends to the Bayesian case by considering
  the discrepancy of the expected interpretation of the auditor; this
  is defined precisely below. It is convenient to generalize the
  notion to yield a default result appropriate for cases
  when no ballot matches the identifier declared in a CVR; this is the $\bot$ case in the definition below.
\begin{definition}[Bayesian discrepancy] Let
  $E=(\candidates, \mathbf{B}, \size)$ be an election
  and let $\cvr = ((\iota_1,\pred_1),\ldots,(\iota_t,\pred_{\size_\cvr}))$ be a
  Bayesian CVR with declared winner $\cW$.  For a ballot
  $\mathbf{b} \in \mathbf{B}$ and a distribution
  $\pred: \interpretations \rightarrow \R$, the \emph{discrepancy} is
  \[
    \disc[\pred, \mathbf{b}; \cW] = \max_{\cA \in \candidates
      \setminus \cW} \left(\bigl(\cinterpretation^{\vaverage}(\cW) -
        \cinterpretation^{\vaverage}(\cA) \bigr)-\bigl(
        \tinterpretation^{\vaverage}_{\mathbf{b}}(\cW) -
        \tinterpretation^{\vaverage}_{\mathbf{b}}(\cA)\bigr)\right)
\]
where, as above, $\cinterpretation^{\vaverage}(\cA) = \Exp[I(\cA)]$
with $I$ distributed according to $\pred$.  We expand this definition
to apply, additionally, to a special symbol $\mathbf{\bot}$:
\[
  \disc[\pred, \perp ; \cW] = \max_{\cA \in \candidates \setminus
    \cW} \left(\cinterpretation^{\vaverage}(\cW) -
    \cinterpretation^{\vaverage}(\cA) \right) + 1\,.
\]
For an identifier $\iota$ define
\[
  \disc_{\cvr, \iota} =
  \begin{cases}
    \min \;\bigl\{ \disc[\pred_{\iota}, \mathbf{b}; \cW] \mid
    \mathbf{b} \in \mathbf{B}_\iota\bigr\} & \text{if $\mathbf{B}_{\iota} \neq \emptyset$,}\\
    \disc[\pred_{\iota}, \bot; \cW] & \text{otherwise.}
  \end{cases}
\]
The discrepancy of the cast vote record table $\cvr$ is:
$
  \disc_{\cvr} = \sum_{i=1}^{\size_\cvr} \disc_{\cvr, \iota_i}.
$
\end{definition}

We begin by establishing the fundamental relationship between
discrepancy and margin for invalid CVRs.

\begin{lemma} \label{lem:disc works} Let
  $E = (\candidates,\mathbf{B}, \size)$ be an election and let $\cvr$
  be an invalid CVR for $E$ with declared winner $\cW$ and size
  $\size_\cvr = \size$. Then $\disc_\cvr \ge (\mu_{\cvr} + \mu_E) \cdot \size \ge \mu_{\cvr}\cdot \size$.
\end{lemma}

\ifnum\addproofs=0
As a reminder, all proofs are deferred to the online version~\cite{fuller2024decisive}.
\else 
\begin{proof}[Proof of Lemma~\ref{lem:disc works}]
  Let $\Identifiers^R$ denote the subset of identifiers that identify
  least one ballot: that is, $\Identifiers^R = \{ \iota \in
    \Identifiers(\cvr) \mid \mathbf{B}_\iota \neq \emptyset\}$.
  We call these ``represented'' identifiers.  For each
  $\iota \in \Identifiers^R$, let $\mathbf{c}_{\iota}$ be a ballot in
  $\mathbf{B}_\iota$ for which
  $\disc[\pred_\iota,\mathbf{c}_{\iota}; \winner]
  =\disc_{\cvr,\iota}$. As
  $\mathbf{B}_\iota \cap \mathbf{B}_{\iota'} = \emptyset$ for any
  $\iota \neq \iota'$, these chosen ballots $\mathbf{c}_\iota$ are
  distinct. Considering that
  $|\Identifiers(\cvr)| = \size_\cvr = \size = |\mathbf{B}|$, the number of
  ``unassigned'' ballots in $\mathbf{B}$ (that do not appear in
  $\{ \mathbf{c}_{\iota} \mid \iota \in \Identifiers^R\}$) is the same
  as the number of unrepresented identifiers (appearing in $\Identifiers(\cvr)$
  but not in $\Identifiers^R$). By choosing an arbitrary
  one-to-one correspondence between unrepresented identifiers and
  unassigned ballots, we may extend the correspondence
  $\iota \leftrightarrow \mathbf{c}_\iota$ to all identifiers in
  $\Identifiers(\cvr)$ and all ballots in $\mathbf{B}$ so that it is a
  one-to-one correspondence between these two sets.
  To complete the proof, we observe that
  \begin{align}
    \disc_{\cvr} &= \sum_{i=1}^t \disc_{\cvr, \iota_i}\\
                 &= \;\;\sum_{\mathclap{\iota \in \Identifiers^R}} \min \left\{ \disc[\pred_{\iota}, \mathbf{b}; \cW] \mid
                                                        \mathbf{b} \in \mathbf{B}_\iota\right\} + \sum_{\mathclap{\substack{\iota \not\in \Identifiers^R\\(\iota \in \Identifiers(\cvr))}}} \disc[\pred_{\iota}, \bot; \cW] \nonumber
    \\
                                                      &\geq \;\;\sum_{\mathclap{\iota \in \Identifiers^R}} \disc[\pred_{\iota}, \mathbf{c}_{\iota}; \cW]  + \sum_{\mathclap{\substack{\iota \not\in \Identifiers^R\\(\iota \in \Identifiers(\cvr))}}} \disc[\pred_{\iota}, \mathbf{c}_\iota; \cW] \nonumber \\
                                                      &\geq \;\;\sum_{\mathclap{\iota \in \Identifiers}} \disc[\pred_{\iota}, \mathbf{c}_{\iota}; \cW]\,, \label{eq:pickup-here} 
  \end{align}
  where we have used the fact---immediate from the definition---that for any distribution $\mathcal{P}$ and any ballot $\mathbf{b}$,
  \[
    \disc[\mathcal{P},\bot;\winner] \geq
    \disc[\mathcal{P},\mathbf{b};\winner]\,.
  \]
  Let $\cB \neq \cW$ be a candidate for which
  $E^{\vaverage}(\cB) \geq E^{\vaverage}(\cA)$ for all candidates
  $\cA$. (If there is no winner in $E$, such a candidate always exists
  but may ``tie'' with some other candidate(s).) Then, again by the
  definition of discrepancy, for each $\iota$
  \[
    \disc[\mathcal{P}_\iota,\mathbf{c}_{\iota};\winner] \geq
    \left(\bigl(\cinterpretation^{\vaverage}_\iota(\cW) -
      \cinterpretation^{\vaverage}_\iota(\cB) \bigr)-\bigl(
      \tinterpretation^{\vaverage}_{\mathbf{c}_{\iota}}(\cW) -
      \tinterpretation^{\vaverage}_{\mathbf{c}_\iota}(\cB)\bigr)\right)
  \]
  Returning to the expression~\eqref{eq:pickup-here}, we then conclude
  that
  \begin{align*}
    \sum_{\mathclap{\iota \in \Identifiers(\cvr)}} \;\disc[\pred_{\iota}, \mathbf{c}_{\iota}; \cW] &\geq \sum_i \left( \bigl(\cinterpretation^{\vaverage}_\iota(\cW) -
      \cinterpretation^{\vaverage}_\iota(\cB) \bigr) + \bigl(
      \tinterpretation^{\vaverage}_{\mathbf{c}_{\iota}}(\cB) -
    \tinterpretation^{\vaverage}_{\mathbf{c}_\iota}(\cW)\bigr) \right)\\
                                                                                                   &= \sum_i \bigl(\cinterpretation^{\vaverage}_\iota(\cW) -
      \cinterpretation^{\vaverage}_\iota(\cB) \bigr) + \sum_{\mathbf{b} \in \mathbf{B}} \bigl(
      \tinterpretation^{\vaverage}_{\mathbf{b}}(\cB) -
      \tinterpretation^{\vaverage}_{\mathbf{b}}(\cW)\bigr)\\
    &= \bigl(\cvr(\cW) -
      \cvr(\cB)\bigr) + \bigl(E(\cB) -
      E(\cW)\bigr)\\
    &\geq \size \cdot (\mu_{\cvr} + \mu_E)\,.
  \end{align*}
  In the first equality above we use the fact that
  $\iota \leftrightarrow \mathbf{c}_{\iota}$ is a one-to-one
  correspondence, so that each ballot appears once in this sum. We conclude that $\disc_{\cvr} \geq \size(\mu_\cvr + \mu_E)$, as desired.
\end{proof}
\fi

\subsection{Adaptive Statistical Tests}

The standard approach for ballot comparison audits determines an
  experiment that appropriately samples ballots in order to obtain
  discrepancy samples; the procedure concludes with $\inconclusive$
  unless one can statistically reject the hypothesis that discrepancy
  is at least $\mu_{\cvr}$, as this is true for all invalid elections
  (Lemma~\ref{lem:disc works}).  Such hypothesis testing requires a
  one-sided statistical test.  In fact, as we consider environments that need
not respond independently during each ballot request, we require a
slightly richer family of tests, formulated in this setting
by~\cite{harrison2022adaptive}. We briefly lay out the definitions
here. The parameter $\delta$ will ultimately be set to $\mu$ in
  our application.

\begin{definition}~\cite[Definition 8]{harrison2022adaptive}
  A sequence of bounded (real-valued) random variables $X_1, \ldots$
  are said to be \emph{$\delta$-dominating} if, for each $t \geq 0$,
  $
  \Exp[X_t \mid X_1, \ldots, X_{t-1}] \geq \delta\,.
  $
  We apply  this terminology to the distribution
  $\mathcal{D}$ corresponding to the random variables, writing $\delta \unlhd \mathcal{D}$.
\end{definition}

\begin{definition}~\cite[Definition 9]{harrison2022adaptive} Let $\Sigma = [-2, 2]$. A \emph{stopping time} is a function
  $\Stop: \Sigma^* \rightarrow \{0,1\}$ so that for any sequence
  $x_1, x_2, \ldots$ of values in $\Sigma$ there is a finite prefix
  $x_1, \ldots, x_k$ for which $\Stop(x_1, \ldots, x_k) = 1$.
    
  For a sequence of random variables $X_1, \ldots$ taking values in
  $\Sigma$, let $\FirstStop(X_1, \ldots)$ be the random variable given
  by the smallest $t$ for which $\Stop(X_1, \ldots, X_t) = 1$. This
  naturally determines the random variable
  $X_1, \ldots, X_{\FirstStop}$, the prefix of the $X_i$ given by the
  first time $\Stop() = 1$.
  \label{def:stop time}
\end{definition}

\begin{definition}\label{def:audit-test}~\cite[Definition 10]{harrison2022adaptive}
  An \emph{adaptive audit test}, denoted $(\Stop, \Criterion)$, is
  described by two families of functions, $\Stopdelta$ and
  $\Criteriondelta$. For each $0 < \delta \le 1$,
  \begin{enumerate}
  \item $\Stopdelta$ is a stopping time, as in Definition~\ref{def:stop time}, and 
  \item $\Criteriondelta:\Sigma^* \rightarrow \{0,1\}$ is
    the \emph{rejection criterion}. 
  \end{enumerate}
  Let $\mathcal{D}$ be a probability distribution on $\Sigma^\N$; for
  such a distribution, define
  $\alpha_{\delta,{\mathcal{D}}} = \Exp[\Criteriondelta(X_1, \ldots,
  X_{\tau})]$ where $X_1, \ldots$ are random variables
  distributed according to $\mathcal{D}$ and $\tau$ is determined by
  $\Stopdelta$.
The \emph{risk} of the test $(\Stop, \Criterion)$ is
  \begin{equation}
  \label{def:audit test}
    \alpha = \sup_{\substack{0 < \delta \le 2\\ \delta \unlhd \mathcal{D}}} \alpha_{\delta,{\mathcal{D}}}\,.
  \end{equation}
\end{definition}

\section{Bayesian Comparison Audits; Leveraging Marginal Marks}
\label{sec:questionable audits}
We present our auditor and analysis using a cryptographic game between the \emph{auditor} and the
\emph{environment}, in the spirit of~\cite{harrison2022adaptive}. The auditor naturally reflects the role played by
an auditing team in a conventional audit; as such it is provided with
a CVR, the actual size of the election (in real-world settings this is
provided by a trusted \emph{ballot manifest}), and the ability to
request ballots by identifier in order to carry out the audit. When
the auditor has completed the audit, it returns either the token
$\inconclusive$ or $\consistent$. The environment, on the other hand,
is responsible for providing physical ballots when they are
requested. Note, in particular, that the auditor's access to ballots
is entirely mediated by the environment, which may choose not to
return ballots when they are requested, make choices about which
ballot to return when multiple ballots share an identifier, etc. The
power of this modeling framework is that the critical concept of risk
can be defined \emph{in the worst case over all possible behaviors of
  the environment}. This reflects a wide variety of
failures---inadvertent or malicious---that occur in practical
audits. 

The formal Auditor-Environment game is described in
Figure~\ref{fig:questionable auditing game}. We remark that each
ballot ``delivered'' by the environment in the game generates a single
draw from the associated ground truth distribution of interpretations;
multiple requests for the same ballot are permitted by the convention,
which generate new independent samples from the ground truth
distribution (modeling a situation where an audit board evaluates the
ballot afresh each time it is sampled).

\myparab{The formal auditor.} The auditor we consider is a standard
``comparison auditor''~\cite{lindeman2012gentle} adapted to the setting where discrepancy is
computed with respect to the expected value of the predicted
distribution in the CVR rather than a single
interpretation. Fortunately, once discrepancy is redefined in this
way, the remaining analysis of the standard auditor requires no
significant changes. 

\begin{figure}[t]
\small
  \begin{framed}
  Auditor ($\Auditor$)--Environment ($\Environment$) game for 
    election $E = (\mathbf{B}, \candidates, \size)$ and CVR $\cvr$
    \begin{enumerate}[noitemsep]
  \item \textbf{Setup}.
    \begin{enumerate}[noitemsep]
    \item\textbf{Ballot and tabulation delivery (to $\Environment$).} The
      physical ballots $\mathbf{B}$ is given to $\Environment$.
    \item\textbf{Ballot manifest and CVR delivery
        (to $\Auditor$).}  The size 
      $\size$ and the $\cvr$ are
      given to the auditor $\Auditor$.
    \end{enumerate}
  \item \textbf{Audit}. $\Auditor$ repeatedly makes a ballot request to $\Environment$, or chooses to conclude the
    audit:
    
    \textbf{A ballot request}. $\Auditor$ requests a ballot
      from $\Environment$ with identifier
      $\iota \in \{0,1\}^*$.
      
 $\Environment$ does one of two things:
 \begin{enumerate}
 \item Responds with a ballot $\mathbf{b}$ with identifier $\iota'$;
   in this case, $\Auditor$ is given $\iota'$ and $I$, an
   interpretation drawn from $\truth_{\mathbf{b}}$.
       \item Responds with \textbf{No
      ballot}; this is forwarded to $\Auditor$.
      \end{enumerate}
      

    \item \textbf{Conclusion}. $\Auditor$ returns one of the two values:
$\consistent$ or
$\inconclusive$.
\end{enumerate}
            \vspace{-.1in}
\end{framed}
            \vspace{-.15in}
\caption{The $\RLA_{\Auditor,\Environment}(E, \cvr)$ auditing game for comparison audits with multiple interpretations.}
            \vspace{-.1in}
\label{fig:questionable auditing game}
\end{figure}

The formal auditor for the Bayesian setting is presented in
Figure~\ref{fig:questionable-auditor}.
    \begin{figure}[t]
    \small
  \begin{framed}
    \underline{$\Auditor[\Stop, \Criterion](E = (\candidates, \mathbf{B}, \size), \cvr)$}
    \begin{enumerate}[noitemsep]
    \item Receive $\cvr = \bigl((\iota_1,\pred_1), \ldots, (\iota_t,\pred_{\size_\cvr})\bigr)$.
\item 
    If $\cvr$ has repeated identifiers or $\size \neq \size_\cvr$, return
    $\inconclusive$.
  \item If $\cvr$ declares a winner $\cW$, define $\mu:=\mu_{\cvr}$;\\
    otherwise, return $\inconclusive$. \label{step:compute margin}
    \item Initialize $\iter=0$.  
    \item Repeat
      \begin{enumerate}[leftmargin=1cm,nosep]
      \item Increment $\iter := \iter + 1$.
      \item Perform $\disc_{\iter} := \mathtt{BasicExperiment}$
      \end{enumerate}
      until $\Stopmu(\disc_1,..., \disc_\iter)=1$
    \item If $\Criterionmu(\disc_1,..., \disc_\iter)=1$ return $\mathtt{Consistent}$; otherwise return $\inconclusive$.
    \end{enumerate}
    
\underline{$\mathtt{BasicExperiment}$}:
    \begin{enumerate}[noitemsep]
    \item $\mathtt{RowSelect}$: Select a row $r \in [\size]$ uniformly.
    \item       Let $\iota$ be the identifier in row $r$;\label{step:start ballot}
  request delivery of $\iota$.  
\item If a ballot $\mathbf{b}$ was delivered with
  identifier $\iota'=\iota$ and interpretation $I^*$,  return 
  \[\act := \max_{\cA\in \candidates \setminus \cW} \left(\cinterpretation^{\vaverage}_r(\cW) - \cinterpretation^{\vaverage}_r(\cA)  - \left( I^*(\cW) - I^*(\cA)\right)\right).\] 
Else, return  \label{step:ballot id check}
  $\displaystyle\act := \max_{\cA\in \candidates \setminus \cW} \left(\cinterpretation^{\vaverage}_r(\cW) - \cinterpretation^{\vaverage}_r(\cA) \right) + 1$. 
    \label{step:stop ballot}
\end{enumerate}
\vspace{-.1in}
\end{framed}
\vspace{-.1in}
\caption{The auditor $\Auditor[(\Stop, \Criterion)]$.}
            \vspace{-.1in}
\label{fig:questionable-auditor}
\end{figure}
We now define the risk of a Bayesian audit: this is an upper
  bound on the probability
  (taken over all elections and environments) that the audit outputs
  $\consistent$ if the CVR is in fact invalid.

\begin{definition}[Risk of Bayesian audit]
\label{def:valid environment}
  Let $\Auditor$ be an Auditor.  For election $E$ and environment
  $\Environment$ let
  $\RLA_{\Auditor,\Environment}(E)$ denote the random variable
  equal to the conclusion of the audit as described in Figure~\ref{fig:questionable auditing game}.
An auditor $\Auditor$ has \emph{$\risk$-risk} (or
    \emph{$\risk$-soundness}) if, for all elections $E$, all invalid $\cvr$s and all environments $\Environment$,    \label{def:risk}
\[
      \Pr_{\Auditor, \Environment}[\RLA_{\Auditor,\Environment}(E, \cvr) = \consistent] \leq \risk\,.
    \]
  \end{definition}

\begin{theorem}
\label{thm:quest audits}
Let $E$ be an election and $\cvr$ be an invalid CVR. For any
environment $\Environment$, let $\disc^{\obs}_{i,\Environment}$ denote
the random variable computed in $\mathtt{BasicExperiment}$ during
iteration $i$; then
$\mathbb{E}[\disc^{\obs}_{i, \Environment}] \ge \disc_\cvr/\size \ge
\mu_{\cvr}$.
    
Let $(\Stop, \Criterion)$ be an adaptive audit test with risk
$\alpha$. Let $\Auditor$ be as in
Figure~\ref{fig:questionable-auditor}, 
$\Auditor[\Stop, \Criterion]$ has risk $\alpha$.
\end{theorem}    
\ifnum\addproofs=1
\begin{proof}
The proof of Theorem~\ref{thm:quest audits} naturally follows from Lemma~\ref{lem:disc works} after noting that for all values $x\in \Sigma, i \in\mathbb{N}$ it must be the case that $\Pr[\disc^{\obs}_{i,\Environment}\le x] \le \Pr_{i\leftarrow [1,t]}\left[\disc_{\cvr, \iota_i}\le x\right]$.  Thus, it follows that $\Exp[\disc^{\obs}_{i,\Environment}] = \disc_{\cvr}/\size \ge \mu_{\cvr}$.
\end{proof}
\fi

\ifnum\conference=1
\subsection{Conservative Comparison Audits}
\else
\section{Conservative Comparison Audits}
\fi
\label{sec:conservative}

Our second approach is called a \emph{conservative comparison audit}.
At a high level, this can be seen as replacing the probability
distribution from the previous section with a set of possible
interpretations.  For a declared winner $\cW$, the auditor runs the audit as
though the interpretation that minimizes $\mu_{\cW}$ occurs with
probability $1$.  When accurate prediction of the auditor distribution
is possible this model has worse results than Bayesian interpretation
but doesn't suffer from the poor efficiency when one overshoots the
probability of the auditor (see results in the next subsection). As a
second advantage, for our modeling in the conservative setting, the
environment is allowed to choose the interpretation of the ballot
selected by the audit board among the possible interpretations of the
ballot.  This is opposed to the previous section where this
interpretation was i.i.d.\ sampled by the game. 

We continue to use the common definitions and terminology for
  ballots and interpretations. The major change is rather than calling
  for a ground truth distribution $\truth_{\mathbf{b}}$, we merely
  define a set $\struth_{\mathbf{b}}$ of \emph{possible
    interpretations} for this ballot. (That is, interpretations that
  could reasonably be the conclusion of an audit board's inspection of
  the ballot.) In the setting with a ground-truth distribution, the
  set $\struth_{\mathbf{b}}$ would naturally correspond to the
  \emph{support} of the distribution
  $\struth_{\mathbf{b}} = \{ I \in \interpretations \mid
  \truth_{\mathbf{b}}(I) > 0\}$.

\begin{definition}[Conservative Interpretation limits]
  \label{def:conservative election winner}
  Let $E = (\candidates, \mathbf{B}, \size)$ be an election.  For a
  ballot $\mathbf{b} \in \mathbf{B}$ and a candidate
  $\cA \in \candidates$, define
  \begin{align*}
    \tinterpretation_{\mathbf{b}}^{\vfloor}(\cA) &= \min \{ I(\cA) \mid I \in \struth_{\mathbf{b}} \}\,,\\
    \tinterpretation_{\mathbf{b}}^{\vceil}(\cA) &= \max \{ I(\cA) \mid I \in \struth_{\mathbf{b}} \}\,.
  \end{align*}
  Thus $\tinterpretation_{\mathbf{b}}^{\vceil}(\cA)$ and
  $\tinterpretation_{\mathbf{b}}^{\vfloor}(\cA)$ indicate the most and
  least favorable interpretations of the ballot for the candidate
  $\cA$.
  We additionally
  define
\[
  E^\vfloor(\cA) = \sum_{\mathbf{b} \in \mathbf{B}}
  \tinterpretation_{\mathbf{b}}^\vfloor(A)\,, 
\qquad\text{and}\qquad
  E^\vceil(\cA) = \sum_{\mathbf{b} \in \mathbf{B}}
  \tinterpretation_{\mathbf{b}}^\vceil(A)\,.
\]
\end{definition}

\begin{definition}[Conservative margin and winner]
  The \emph{conservative margin of $\cA$ with respect to $\cB$} 
  compares the worst and best case interpretations for
  two candidates:
\[
  \mu_{\cA, \cB}^{\sure} = \frac{E^{\vfloor}(\cA) -  E^{\vceil}(\cB)}{\size}.
\]
We then define
$\mu_{\cA}^{\sure} = \min_{\cB \in \candidates\setminus \{\cA\}}
\mu_{\cA, \cB}^{\sure}$. If there exists some candidate $\cW$ such
that $\mu_{\cW}^{\sure} >0$ we call this candidate the (conservative)
winner and denote this quantity by $\mu^{\sure}$.  If a conservative
winner exists, then it is unique.  If there is no conservative winner,
the election is called \emph{indeterminate}.  We call a candidate
$\loser$ a conservative loser if there exists some other candidate
$\cA$ such that $\mu_{\loser, \cA}^{\sure} <0$.
\end{definition}

Observe that losing candidates may exist even in indeterminate
  elections and that, in general, a losing candidate candidate cannot
  tie the election even if every ballot is given the interpretation
  most favorable to the candidate. As mentioned above, conservative
CVRs specify a set of declared interpretations for each ballot.

  \begin{definition}[Conservative Cast-Vote Record Table (CVR)]
  \label{def:conservative cvr}
  Let $\candidates$ be a set of candidates. A \emph{conservative
    Cast-Vote Record table (CVR)} is a sequence of pairs
  \[
    \cvr^{\sure} = \bigl((\iota_1,\spred_1), \ldots, (\iota_t,\spred_{\size_\cvr})\bigr),
  \]
  where the $\iota_r$ are distinct bitstrings in $\{0,1\}^*$ and each
  $\spred_r \subset \interpretations$ is a subset of interpretations.
  The CVR declares that for each row $r$ the
  interpretation of the ballot labeled $\iota_r$ lies in the set
  $\spred_{r}$.  
  For each candidate $\cA \in \candidates$, we define
  \begin{align*}    \cinterpretation^{\vfloor}_r(\cA) &= \min \{ I(A) \mid I \in \spred_r\},  \\\cinterpretation^{\vceil}_r(\cA) &= \max \{ I(A) \mid I \in \spred_r\},
  \end{align*}
  \[
    \cvr^{\vfloor}(\cA) = \sum_{r=1}^{\size_\cvr} \cinterpretation_r^{\vfloor}(\cA)\,, \qquad \text{and}\qquad
    \cvr^{\vceil}(\cA) = \sum_{r=1}^{\size_\cvr} \cinterpretation_r^{\vceil}(\cA)\,.
  \]
\end{definition}

\begin{definition}[Declared winners, losers, and contradictory CVRs.] If there is a
 candidate $\cA \in \candidates$ with the property that
 $\cvr^{\vfloor}(\cA) > \cvr^{\vceil}(\cB)$ for every
 $\cB \in \candidates \setminus \{\cA\}$, then we say that $\cA$ is
 the \emph{declared conservative winner} according to $\cvr^{\sure}$.
 If there is no such candidate, we say that $\cvr^{\sure}$ is
 \emph{indeterminate}.  Any candidate $\loser \in \candidates$ for
 which $\cvr^{\vceil}(\loser) < \cvr^{\vfloor}(\cA)$ for some
 candidate $\cA \in \candidates$ is a \emph{declared conservative
   loser}. We adopt the same naming conventions for such conservative
 CVRs as we do with Bayesian CVRs (that is, for the set of
 identifiers, size, and treating identifiers as indices for $P_r$ and
 $\cinterpretation_r$).

 Finally, we say that two CVRs are \emph{contradictory} if there is a candidate declared as a winner in one and a loser in the other.
\end{definition}

    An election or CVR can be indeterminate without a tie: e.g., there may be two candidates $\cA$ and $\cB$ for which $\cvr^{\vceil}(\cA) > \cvr^{\vfloor}(\cB)$ and $\cvr^{\vceil}(\cB) > \cvr^{\vfloor}(\cA)$.

\begin{definition}[Conservative Declared Margin]
We call a candidate $\loser$ a conservative loser if there exists some other candidate $\cA$ such that $\mu_{\cA,\loser}^{\sure} >0$.  
  The
  \emph{declared margin} of a conservative $\cvr$ with \emph{declared winner} $\cW$ is the quantity
    \[
      \mu_{\cvr}^{\sure} := \min_{\cA\in \candidates \setminus \cW}
      \frac{\cvr^{\vfloor}(\cW) - \cvr^{\vceil}(\cA)}{\size}.
    \]
    \end{definition}

\begin{definition}[Conservative discrepancy] Let
  $E=(\candidates, \mathbf{B}, \size)$ be an election
  and let $\cvr = ((\iota_1,\spred_1),\ldots,(\iota_t,\spred_t))$ be a
  conservative CVR with declared winner $\cW$.  For a ballot
  $\mathbf{b} \in \mathbf{B}$ and interpretation $I$, the
  \emph{conservative discrepancy} of $I$ with respect to $\mathbf{b}$ is
  \[
    \disc^{\sure}[I, \mathbf{b}; \cW] =
    \max_{\cA \in \candidates \setminus
      \cW} \left( \bigl(I(\cW) - I(\cA)\bigr)- \bigl(\struth_{\mathbf{b}}^{\vceil}(\cW) - \struth_{\mathbf{b}}^{\vfloor}(\cA))\right)\,.
  \]
  We expand this definition to apply, additionally, to a special symbol $\mathbf{\bot}$:
  \[
    \disc^{\sure}[I,\perp ; \cW]  =  \max_{\cA \in \candidates \setminus \cW} \bigl(I(\cW) - I(\cA)\bigr) + 1\,.
  \]
  The discrepancy of a cast vote record table $\cvr$ is:
  \[
    \disc_{\cvr}^{\sure} = \sum_{i=1}^t \min \bigl\{ \disc[I, \mathbf{b}; \cW]
    \;\bigm|\; I \in \mathtt{I}_i^{\cvr}, \text{${\mathbf{b}} = \mathbf{\bot}$ or $\id_{\mathbf{b}} =
      \iota_i$}\bigr\}\,.
  \]
\end{definition}

\begin{definition}[Conservative consistency and validity]
  A
  $\cvr^+ = \bigl((\iota_1,\spred_1), \ldots, (\iota_t,\spred_t)\bigr)$
  is \emph{consistent} with $E$ if the ballots $\mathbf{B}$ are
  uniquely labeled, the identifiers appearing in $\cvr$ are identical
  to those appearing on the ballots of $E$, and for each ballot
  $\mathbf{b} \in \mathbf{B}$,
  $\struth_{\mathbf{b}} \subset \spred_{\iota_\mathbf{b}}.$ When $\struth_{\mathbf{b}} = \spred_{\id_\mathbf{b}}$
  for each ballot, we say that $\cvr$ is the \emph{canonical} CVR for
  the election $E$. We only define these notions for uniquely labeled
  elections.
  
  For election $E$ with
  winner $\cW$, a $\cvr$ is invalid if it declares a winner other than
  $\cW$.
\end{definition}

\ifnum\conference=1
\subsubsection{Conservative Game and Auditor}
\else
\subsection{Conservative Game and Auditor}
\fi

We consider the following adaption of the auditing game where the environment is allowed to specify the interpretation returned to the auditor in Figure~\ref{fig:questionable auditing game min} with the change in boldface.  The auditor is identical to that in Figure~\ref{fig:questionable-auditor} except the Bayesian $\cvr$ margin in Step~\ref{step:compute margin} is replaced with $\mu_{\cvr}^{\sure}$.  

\begin{figure}[t]
\small
  \begin{framed}
  Auditor ($\Auditor$)--Environment ($\Environment$) game for 
    election $E = (\mathbf{B}, \candidates, \size)$ and CVR $\cvr$
    \begin{enumerate}[noitemsep]
  \item \textbf{Setup}.
    \begin{enumerate}[noitemsep]
    \item\textbf{Ballot and tabulation delivery (to $\Environment$).} The
      physical ballots $\mathbf{B}$ is given to $\Environment$.
    \item\textbf{Ballot manifest and CVR delivery
        (to $\Auditor$).}  The size 
      $\size$ and the $\cvr$ are
      given to the auditor $\Auditor$.
    \end{enumerate}
  \item \textbf{Audit}. $\Auditor$ repeatedly makes a ballot request to $\Environment$, or chooses to conclude the
    audit:
    
    \textbf{A ballot request}. $\Auditor$ requests a ballot
      from $\Environment$ with identifier
      $\iota \in \{0,1\}^*$.
      
 $\Environment$ either responds a ballot $\mathbf{b}$ with single identifier $\iota'$ and \textbf{an interpretation $I \in \struth_{\mathbf{b}}$} or responds with \textbf{No
      ballot.}

  \item \textbf{Conclusion}. $\Auditor$ returns one of the two values:
$\consistent$ or
$\inconclusive$.
\end{enumerate}
\vspace{-.15in}
\end{framed}
\vspace{-.15in}
\caption{The $\RLA_{\Auditor,\Environment}(E, \cvr)$ auditing game for comparison audits with multiple interpretations.}
\vspace{-.15in}
\label{fig:questionable auditing game min}
\end{figure}
\begin{theorem}
\label{thm:quest audits min}
Let $E$ be an election and $\cvr$ be an invalid CVR. For any environment $\Environment$ engaging in Figure~\ref{fig:questionable auditing game min}, let $\disc^{\obs}_{i,\Environment}$ denote the random variable computed in $\mathtt{BasicExperiment}$ then 
$\mathbb{E}[\disc^{\obs}_{i, \Environment}] \ge \disc_\cvr/\size \ge \mu.
$
    where $\mu$ is the value computed in Step~\ref{step:compute margin}.
    
Let $(\Stop, \Criterion)$ be an adaptive audit test with risk
$\alpha$. Let $\Auditor$ be as in
Figure~\ref{fig:questionable-auditor} with $\mu_{\cvr}$ replaced with $\mu_{\cvr}^{\sure}$, 
$\Auditor[\Stop, \Criterion]$ has risk $\alpha$.
\end{theorem}    
\ifnum\addproofs=1
\begin{proof}
The proof of Theorem~\ref{thm:quest audits min} proceeds similarly to Theorem~\ref{thm:quest audits}.  Here we provide a substitute for Lemma~\ref{lem:disc works} for the conservative setting.

\begin{lemma} \label{lem:disc works min} Let
  $E = (\candidates,\mathbf{B}, \size)$ be an election and let $\cvr$
  be an invalid CVR for $E$ with declared winner $\cW$ and size
  $t = \size$. Then $\disc_\cvr^{\sure} \ge (\mu_{\cvr}^{\sure} + \mu_E^{\sure}) \cdot \size \ge (\mu_{\cvr}^{\sure})\cdot \size$.
\end{lemma}
\begin{proof}
We focus on the setting when $\mathbf{B}$ is uniquely labeled and all identifiers appear in the uniquely labeled $\cvr$.  See the proof of Lemma~\ref{lem:disc works} for why this setting suffices. 

Let $\cW_{\cvr}$ be the declared conservative winner and let $\cW_{E} \neq \cW_{\cvr}$ the conservative winner of the election.
By definition, $E^\vfloor(\cW_E)> E^\vceil(\cW_{\cvr})$.

By the
  definition of discrepancy, for each $\iota$
  \[
    \disc^{\sure}[\mathcal{P}_\iota,\mathbf{c}_{\iota};\winner_{\cvr}] \geq
    \left(\bigl(\cinterpretation^{\vfloor}_\iota(\cW_{\cvr}) -
      \cinterpretation^{\vceil}_\iota(\cW_E) \bigr)- \bigl(
      \tinterpretation^{\vfloor}_{\mathbf{c}_{\iota}}(\cW_{\cvr}) -
      \tinterpretation^{\vceil}_{\mathbf{c}_\iota}(\cW_E)\bigr)\right)
  \]
We then conclude
  that
  \begin{align*}
       \disc_{\cvr}^{\sure} &\ge  \sum_{\iota \in \Identifiers(\cvr)} \disc[\pred_{\iota}, \mathbf{c}_{\iota}; \cW] \\&\geq \sum_i \left(\bigl(\cinterpretation^{\vfloor}_\iota(\cW_{\cvr}) -
      \cinterpretation^{\vceil}_\iota(\cW_E) \bigr) + \bigl(
      \tinterpretation^{\vfloor}_{\mathbf{c}_{\iota}}(\cW_{E}) -
      \tinterpretation^{\vceil}_{\mathbf{c}_\iota}(\cW_{\cvr})\bigr)\right)\\
    &= \sum_i \bigl(\cinterpretation^{\vfloor}_\iota(\cW_{\cvr}) -
      \cinterpretation^{\vceil}_\iota(\cW_{E}) \bigr) + \sum_{\mathbf{b} \in \mathbf{B}} \bigl(
      \tinterpretation^{\vfloor}_{\mathbf{b}}(\cW_E) -
      \tinterpretation^{\vceil}_{\mathbf{b}}(\cW_{\cvr})\bigr)\\
    &= \bigl(\cvr(\cW_{\cvr}) -
      \cvr(\cW_E)\bigr) + \bigl(E(\cW_e) -
      E(\cW_{\cvr})\bigr)\\
    &\geq \size \cdot (\mu_{\cvr}^{\sure} + \mu_E^{\sure})\,.
  \end{align*}
  In the first equality above we use the fact that
  $\iota \leftrightarrow \mathbf{c}_{\iota}$ is a one-to-one
  correspondence, so that each ballot appears once in this sum. We conclude that $\disc_{\cvr} \geq \size(\mu_\cvr^{\sure} + \mu_E^{\sure})$, as desired.
This completes the proof of Lemma~\ref{lem:disc works min}.
\end{proof}
\noindent 
This completes the proof of Theorem~\ref{thm:quest audits min}.
\end{proof}
\fi

\section{Completeness of Bayesian and Conservative Audits; Efficiency Analysis}
\label{sec:efficiency}

The previous subsection established that one can recover a natural RLA
from CVRs that declare multiple interpretations for ballots.  In the
case when all ballots and CVRs have a single interpretation the audit
is a traditional ballot comparison audit.  We also note that the
auditor receives a single interpretation of the ballot.  This models
the fact that an audit board has to produce a final adjudication of a
ballot.  As described above, the purpose of the test is to decide
whether $\mathbb{E} [\disc_{\cvr}/\size] \ge \mu_{\cvr}$.  

Perhaps the most widely deployed test meeting the demands outlined in
the previous section is the Kaplan-Markov test (though there are other
natural choices~\cite{Stark:Conservative,stark2020sets,waudby2021rilacs,spertus2023cobra}); in particular, this is the test featured in
the \href{https://www.voting.works/risk-limiting-audits}{Arlo RLA administration system}. The basic
Kaplan-Markov test (omitting minimum and maximum sample sizes) is as
follows: for a parameter $\gamma>1$, define the value
  \[ \Riskdelta^{(\gamma)}(\disc_1,..., \disc_\ell ) =
    \prod_{\iter=1}^\ell
    \left(\frac{1-\frac{\delta}{2\gamma}}{1-\frac{\disc_\iter}{2\gamma}}\right)\,.
\]
Intuitively, the ``inflation'' parameter $\gamma$ determines the
  relationship between the rate of ballot comparison inconsistencies
  and sample sizes. Common values in the literature are approximately
  1.1, see
  \href{https://bcn.boulder.co.us/~neal/electionaudits/rlacalc.html}{RLACalc}.
Moreover, the stopping time for risk limit $\alpha$ is determined by
the test $\Riskdelta \stackrel{?}{\le} \alpha$;
when this occurs, the test rejects the hypothesis (that the mean is
larger than $\mu$) and, in our setting, the auditor would output
$\consistent$.

\myparab{Simulation parameters.}
We consider a two candidate election with $100,000$ ballots with
$m = .5\%$ marginal ballots and underlying error rates of $o_1=.1\%$,
$u_1=.1\%$, $o_2=.01\%$, and $u_2=.01\%$~\cite{lindeman2012gentle};
Errors $o_1$ and $o_2$ represent ballots with discrepancies of $+1$
and $+2$ respectively. If the CVR lists a vote for the winner $\cW$
and the actual ballot shows a vote for a loser this is an $o_2$ error.
The actual ballot showing no vote results in an $o_1$, this is called
an undervote. The errors $u_1$ and $u_2$ are defined similarly for
discrepancies of $-1, -2$.  For marginal ballots, we assume there are
two possible interpretations: an interpretation for $\cW$ and an
interpretation of an undervote. Throughout the simulations, we
  assign a discrepancy of $1$ and $0$ with equal probability in these
  cases. For the math, the important thing is the distribution of discrepancy, not the votes appearing on the ballots. We compare three approaches in Table~\ref{tab:km sizes}:

\begin{enumerate}
\item In the \emph{Baseline} approach, half of the marginal ballots are counted as votes for the reported winner on the CVR.  When a marginal ballot is selected for audit it is determined as a vote for $\cW$ with probability $50\%$.  Note this can result in discrepancy values of $\{-1,0,1\}$.
\item In the \emph{Bayesian} approach, the margin receives $\cinterpretation^{\vaverage}_r(\cW) = .5$ ``votes'' for each marginal ballot.  As above, when a marginal ballot is selected for audit it is determined as a vote for $\cW$ with probability $50\%$.  However, since $\cinterpretation^{\vaverage}_r(\cA)=.5$ the possible discrepancy values are $\{-.5,.5\}$
\item In the \emph{Conservative} approach, no marginal ballots are included in the margin and only negative discrepancy values are possible if the auditor interprets the ballot as a vote for $\cW$. 
\end{enumerate}

All results perform $5000$ simulations for each parameter setting.  A
simulation pulls a uniform ballot until the risk limit is met.  Each
time a marginal ballot is pulled its interpretation is independently
sampled ($50\%$ probability of vote for $\cW$ and $50\%$ of
blank). RLAs require the most work for small margins which is where
our improvement is evident.  We report on the mean, standard
deviation, median, and $95\%$ threshold of ballots retrieved in
Table~\ref{tab:km sizes}.  In practice, the $95\%$ mark is most
relevant; this reflects the fact that the work to conduct an
additional round of an RLA is often prohibitive---requiring a
distributed process across a whole region---so termination within the
specified sample size with high probability is desirable.  The
Bayesian approach reduces the $95\%$ percentile of ballots sampled by
roughly $10.5\%$.  Roughly, this improvement arises because the
variance of discrepancy for ``marginal'' ballots is reduced from $1/2$
using the baseline approach to $1/4$ using the Bayesian approach.

\begin{table*}[t]

\begin{center}
\footnotesize
\begin{tabular}{ |r|r|r|r|r|| r | r| r| r|| r | r| r| r|} 
\hline
& \multicolumn{4}{c||}{Baseline} & \multicolumn{4}{c||}{Conservative} & \multicolumn{4}{c|}{Bayesian}\\
$\mu$ &	Mean & Stdev & Median & 95\% &Mean & Stdev & Median & 95\%&Mean & Stdev & Median & 95\%\\
\hline
.01&	608 & 210 & 567 & 1028 & 595 & 181 & 576 & 938 & 583 & 175 & 545 & 920\\
\hline
.02&	316 & 78 & 292 & 469 & 314 & 69 & 294 & 420 & 308 & 64 & 292 & 415\\
\hline
.03&	213 & 44 & 202 & 283 & 212 & 38 & 219 & 263 & 210 & 39 & 202 & 271\\
\hline
\end{tabular}
\end{center}
\vspace{-.10in}
\caption{Number of ballots for Kaplan-Markov comparison audit between the baseline and the Bayesian auditor, across margins, and probability $.5$ of CVR and auditor determining marginal mark as for winner, $\alpha=.05, \gamma=1.1$.}
\label{tab:km sizes}
\end{table*}

\begin{table*}[tb]

\begin{center}
\footnotesize
\begin{tabular}{ |r|r|r|r|r|| r | r| r| r|| r | r| r| r|} 
\hline
& \multicolumn{4}{c||}{Baseline} & \multicolumn{4}{c||}{Conservative} & \multicolumn{4}{c|}{Bayesian}\\
$p_m$ &	Mean & Stdev & Median & 95\% &Mean & Stdev & Median & 95\%&Mean & Stdev & Median & 95\%\\
\hline
1&	471 & 115 & 438 & 704 & 514 & 164 & 494 & 824 & 468 & 116 & 438 & 704\\
\hline
.9&	493 & 131 & 454 & 729 & 526&161&494&842 & 490 & 126 & 468 & 738\\
\hline
.8&	520 & 149 & 470 & 817 & 533  & 158&494&824 & 510 & 135 & 485 & 773\\
\hline
.7&	547 & 170 & 502 & 881 &599 & 167&538&879 & 534 & 150 & 504 & 830\\
\hline
.6&	575 & 188 & 539 & 930 &579 & 178 & 544& 944& 559 & 155 & 533 & 856\\
\hline
.5&	608 & 210 & 567 & 1028 & 595 & 181 & 576 & 938 & 583 & 175 & 545 & 920\\
\hline
.4& 630&217 &591 &1071 	& 617 & 181 & 576 & 975 & 616 & 189 & 576 & 1003\\
\hline
.3& 662&231&616&1123& 646 & 199 & 590 & 1020	& 641 & 199 & 596 & 1031\\
\hline
.2&685 & 227&644&1128& 671 & 202 & 627 & 1071 & 671 & 211 & 619 & 1097\\
\hline
.1&707 & 227& 627& 1138& 700 & 215 & 658 & 1108 & 692 & 209 & 630 & 1104\\
\hline
0& 729 & 214 & 658 & 1184 & 729 & 214 & 658 & 1184 & 724 & 215 & 658 & 1184\\
\hline
\end{tabular}
\end{center}
\vspace{-.15in}
\caption{Number of ballots across $p_m$, $\alpha=.05, \gamma=1.1, \mu =.01$.}
            \vspace{-.1in}
\label{tab:km sizes p_m}
\end{table*}

\begin{table*}[t]

\begin{center}
\footnotesize
\begin{tabular}{ |r|r|r|r|r| r | r| r| r| r |r| r| r |r| r| r | } 
\hline
& \multicolumn{3}{c|}{$p_{\Auditor} =p_{\cvr}+.4$ } & \multicolumn{3}{c|}{$p_{\Auditor} =p_{\cvr}+.2$} & \multicolumn{3}{c|}{$p_{\Auditor} =p_{\cvr}$}& \multicolumn{3}{c|}{$p_{\Auditor} =p_{\cvr}-.2$}& \multicolumn{3}{c|}{$p_{\Auditor} =p_{\cvr}-.4$}\\
$p_\cvr$ &Base &  Cons & Bayes &Base &  Cons & Bayes&Base &  Cons & Bayes&Base &  Cons & Bayes&Base &  Cons & Bayes\\
\hline
1&	        &&&&&&704 & 824 &  704 &1053 & 855 & 816 & 1307 & 924 & 750\\
\hline
.9 &       &&&&&&729 & 842 & 738& 1094 & 893 & 840 &1347 & 938 & 763\\
\hline
.8 &       &&&780&810&981&817 & 824 & 773 & 1135 & 938 & 839 & 1347 & 975 & 740\\
\hline
.7&	       &&&856&816&999&881 &879 & 830 & 1139 & 975 & 836 &1373 & 1020 & 768\\
\hline
.6&	        780&810&1129&860&842&1011&930 & 944 & 856  & 1165 & 975 & 850 & 1377 & 1071 & 754\\
\hline
.5 & 790 & 817 & 1164 & 871 & 900 & 1010 & 1028 & 938 & 920 & 1160 & 1052 & 823& 1347 & 1153 & 748 \\
\hline
.4& 816 & 842 & 1203 & 905 & 924 & 1062 & 1071 & 975 & 1003 & 1100 & 1093 & 791 & 1271 & 1184 & 742\\
\hline
.3& 816 & 893 & 1216 & 886 & 960 & 1079 & 1123 & 1020 & 1031 & 1109 & 1108 & 808 &&&\\
\hline
.2&816 & 900 & 1290 & 861 & 975 & 1068 & 1128 & 1071 & 1097 & 1058 & 1184 & 815 &&&\\
\hline
.1& 780 & 938 & 1301 & 820 & 1020 & 1097 & 1138 & 1108 & 1104 & & & &&&\\
\hline
0 & 739 & 1020 & 1347 & 816 & 1102 & 1058 & 1184 & 1184 & 1184 &&& &&&\\
\hline
\end{tabular}
\end{center}
\vspace{-.15in}
\caption{95\% of number of ballots for Kaplan-Markov comparison audit, $\alpha=.05, \gamma=1.1, \mu =.01$.}
            \vspace{-.1in}
\label{tab:km sizes mismatch}
\end{table*}

In the above, both the ground truth (observed by the auditor) and CVR
prediction probability of a marginal mark being interpreted as a mark
is $p_m = .5$.  This is the setting that has the largest improvement
on the variance of discrepancy. However, for varying values of $p_m$
the approach still provides efficiency improvements over the baseline
approach. This is shown in Table~\ref{tab:km sizes p_m}. Note for $p_m\in \{0,1\}$ the baseline and Bayesian approach
are identical as marginal ballots never result in nonzero discrepancy.
In Table~\ref{tab:km sizes mismatch},  we
consider two separate parameters for the probability that the CVR
marks each marginal ballot as a vote for $\cW$, denoted $p_{\cvr}$ and
the probability that the auditor does the same denoted
$p_{\Auditor}$. Interestingly, the baseline approach outperforms the
Bayesian approach when $p_{\cvr}> p_{\Auditor}$ by $\approx 20\%$; however
Bayesian can outperform Baseline by $\approx 30\%$ when
$p_{\cvr}< p_{\Auditor}$.  This is part of the reason we introduce the
conservative approach where one does not have to estimate
probabilities.

\section{Election Contestation and Competitive Ballot Comparison
  Audits}
\label{sec:contested audits}
We consider a new class of post-election audits we call
\emph{competitive audits}. These audits can efficiently distinguish
CVRs that are consistent with the ballots from invalid CVRs and, more
generally, reconcile competing claims of victory made by two
CVRs. This novel auditing technique also provides, as described
earlier, a direct approach to handling election contestation.

To simplify our treatment of these audits, we focus on the
conservative setting, in which CVRs declare, for each ballot, a subset
of possible interpretations without assigning a probability
distribution to the interpretations in the subset.

\myparab{An intuitive survey of the framework.} After an election,
we assume that each candidate $\cA \in \candidates$ is represented by
an \emph{advocate}.  In a preliminary ``rescan phase,'' the advocate
for the candidate $\cA$ may inspect the ballots $\mathbf{B}$ and
submit to the audit a conservative CVR, denoted $\cvr_{\cA}$; the
idea is that the advocate will submit a CVR that is favorable for the
candidate. If the declared conservative winner of $\cvr_{\cA}$ is
$\cA$, we say that this is a ``declaration of victory.''  (Advocates
may also be allowed to announce the presence of duplicate ballot
labels in the election, which can then be corrected by the audit. For
simplicity, however, we simply treat $\mathbf{B}$ as uniquely labeled
in this setting.) Following this, the audit commences with the
``judgment phase:'' after considering some of the ballots, the audit
either declares the audit to be $\inconclusive$ or, for some candidate
$\cA \in \candidates$, declares that ``$\cA$ is the winner.''

\myparab{The judgment phase; reconciling disagreements among the
  CVRs.} If no pair of the submitted CVRs are contradictory, the audit
concludes without considering any ballots. In this non-contradictory
case, if any CVR declares victory the associated candidate is declared
to be the winner of the audit; if, alternatively, no CVR declares a
winner, the audit concludes with $\inconclusive$.

Otherwise, if two submitted CVRs are contradictory, the audit must settle the contradictory claims made by
various submitted CVRs. In this case, some candidate $\cA$ must be a
declared winner according to one CVR and a declared loser according to
another.  We will see that there must then exist an identifier
$\iota^{}$ on which the two CVRs make contradictory
assertions. Intuitively, one of these contradictory CVRs can then be
removed from consideration by fetching the ballot with this identifier, if it exists, and relying on this ballot to settle the
contradictory claims made by the two CVRs.

The natural procedure is complicated by three issues:
\begin{enumerate}
\item two contradictory CVRs may declare different sets of identifiers
  rather than disagreeing on the interpretation of specific ballots;
\item the audit should provide suitable protection against an
  imperfect or malicious environment that may not always respond with
  a given ballot, even when it exists; and
\item the audit should withstand a small collection of errors in
  the creation of CVRs by advocates to reflect the practical
  difficulty of producing a perfect CVR, even by a well-intentioned
  advocate.
\end{enumerate}
We remark that the second issue above forces the audit to treat
``positive evidence'' and ``negative evidence'' differently: for
example, while the delivery of a ballot that disagrees with a
particular CVR, either by possessing an identifier that is not
declared in the CVR or having an interpretation that is inconsistent
with those declared in the CVR, is considered evidence against the
CVR, the failure of the environment to produce a ballot declared in a
particular CVR is not. This cautious convention prevents a malicious
environment from misleading the auditor. An environment that can refuse to 
retrieve ballots can always cause the audit to output $\inconclusive$.

Ultimately, we will prove two facts about the audit.
\begin{itemize}
\item If one of the submitted CVRs is consistent with the ballots and
  ballots are faithfully returned to the auditor during the judgment
  phase, then---except with small probability that can be explicitly
  bounded as a function of the number of samples---the auditor will correctly conclude the audit with ``$\cW$
  is the winner.''  Thus, if the advocate for $\cW$ indeed acts in
  $\cW$'s best interests, $\cW$ can be confident that the audit will
  conclude favorably.
\item If one of the submitted CVRs is consistent with the ballots and an alternate CVR declares that a losing candidate $\loser$ won the election, the audit will not conclude that ``$\loser$ is the winner.'' Furthermore, this guarantee is robust against ballot suppression---it does not require that ballots are correctly returned when requested.
\end{itemize}
Taken together, we see that if a candidate's advocate submits a
consistent CVR, the candidate can be assured a victory if no ballots
are lost or suppressed, and can be assured that no losing candidate is
declared the victor even in the face of ballot suppression.

\subsection{Modeling Competitive Audits}

Formally, we model a competitive audit with two parties, the
\emph{auditor} (denoted $\CAuditor$) and the \emph{environment}
(denoted $\Environment$). The auditor is responsible for analyzing
submitted conservative CVRs, requesting ballots to carry out the audit, and
arriving at a final conclusion. The environment is responsible for
servicing requests by the auditor for individual ballots: in
particular, the environment is in possession of the ballots and, for
each request for a ballot by the auditor, decides which ballot (if
any) will be returned to the auditor and, moreover, how that ballot is
to be interpreted (that is, which possible interpretation is to be adopted).

As mentioned above, this two party modeling of the audit reflects the
fact that even a well-designed auditor may have to contend with such
complexities as loss of ballots, potential malicious suppression of
ballots, and the final choice of an ``auditing board'' in the
interpretation of a ballot. These decisions are in the hands of
the environment.

\begin{definition}[Competitive auditor; competitive environments; honest environments]
  A \emph{competitive auditor} $\CAuditor$ is a randomized procedure
  for carrying out a competitive audit. In the context of an election
  $E = (\candidates, \mathbf{B}, \size)$ with conservative ballots
  $\mathbf{B}$, the auditor is invoked with:
  \begin{enumerate}
  \item the set of candidates $\candidates$, 
  \item the size $\size$
  of the election, and 
  \item a list of conservative CVRs $\cvr_1^{\sure}, \ldots, 
    \cvr_k^{\sure}$.
  \end{enumerate} The audit proceeds in rounds, each round being an
  opportunity for the auditor to request delivery of a ballot (specified by an identifier) along
  with an interpretation of the ballot. After a sequence of requests, the
  auditor concludes the audit, resulting in either the statement
  $\inconclusive$, or ``$\cA$ is the winner'' for some candidate
  $\cA \in \candidates$.

  Ballot requests are handled by a second randomized procedure called
  the \emph{environment}.  The environment is invoked with
  $(\candidates, \mathbf{B}, \cvr_1^{\sure},\ldots, \cvr_k^{\sure})$.
  To each (identifier) request $\iota$ made by the auditor, the
  environment either responds with a ballot
  $\mathbf{b} \in \mathbf{B}$ and a single interpretation
  $I \in \struth_{\mathbf{b}}$ or a ``no ballot'' symbol
  $\mathbf{\bot}$.  While the environment may choose not to return a
  ballot when a matching ballot exists in $\mathbf{B}$, any ballot it
  does return is assumed to match the requested identifier:
  $\iota_\mathbf{b} = \iota$.\footnote{These conventions are a
    convenience: the auditor can check the identifier of a returned
    ballot to ensure that it matches the request and, additionally,
    ensure that the returned interpretation is among those associated
    with the ballot; if either of these verifications fail, the
    auditor treats this ballot response as a $\mathbf{\bot}$.}

An environment is \emph{honest} if it responds to any
  requested identifier with a matching ballot $\mathbf{b}$ and an
  interpretation $I^* \in \mathtt{I}_{\mathbf{b}}$ when such a
  ballot exists.
\end{definition}

\begin{definition}
  Let $E = (\candidates, \mathbf{B}, \size)$ be an election. For a competitive auditor, 
  $\CAuditor$, environment $\Environment$, and  a sequence of conservative CVR tables $\cvr_1^{\sure}, \ldots, \cvr_k^{\sure}$,
  define
  \[
    \CompA[\CAuditor, \Environment; E; \cvr_1^{\sure}, \ldots, \cvr_k^{\sure}]
  \]
  to be the result of the audit (returned by $\CAuditor$) described in
  Figure~\ref{fig:competitive-auditing-game}. As both the auditor and
  the environment may be randomized procedures, this is a random
  variable taking values in the set
  $\candidates \cup \{ \inconclusive\}$.
\end{definition}

\begin{figure}[t]
\small
  \begin{framed}
    Competitive auditor ($\CAuditor$)--Environment ($\Environment$) game
    \[
      \CompA[\CAuditor, \Environment; \mathbf{B}, \candidates, \size]\,.
    \]
    \begin{enumerate}[noitemsep]
    \item \textbf{Advocate CVR generation}. For each candidate
      $\cA \in \candidates$, an advocate for $\cA$---with access to the full
      candidate set $\candidates$ and the physical ballots
      $\mathbf{B}$---generates a cast-vote record table,
      $\cvr_\cA^{\sure}$.
      
    \item \textbf{Setup}.
    \begin{enumerate}[noitemsep]
    \item\textbf{Candidates, ballots, and CVR delivery (to
        $\Environment$).} ($\candidates, \mathbf{B}, \{ \cvr_\cA^{\sure} \mid \cA \in \candidates\})$ are given to the environment
      $\Environment$.
    \item\textbf{Candidate list, ballot manifest, and CVR delivery (to $\CAuditor$).}
$(\candidates,\size, \{ \cvr_\cA^{\sure} \mid \cA \in \candidates\})$ are given to the
      auditor $\CAuditor$.
    \end{enumerate}
  \item \textbf{Audit}. $\CAuditor$ repeatedly makes a request to $\Environment$, or chooses to conclude the
    audit:
    \begin{itemize}[noitemsep]
    \item \textbf{A ballot request}. $\CAuditor$ requests a ballot
      from $\Environment$ with identifier
      $\iota \in \{0,1\}^*$.
      \begin{enumerate}[noitemsep]
      \item $\Environment$ either responds either with an interpretation $I$ where $I \in \struth_\mathbf{b}$ for some  ballot
        $\mathbf{b}$ for which $\iota = \iota_\mathbf{b}$  or responds
        with $\mathbf{\bot}$ (meaning \textbf{No ballot}).
    \end{enumerate}
    \end{itemize}
  \item \textbf{Conclusion}. $\CAuditor$ either returns:
    \begin{description}[noitemsep]
    \item[$\cA$ is the winner] meaning ``The audit is consistent with
      victory by candidate $\cA$,'' or
      \item[$\inconclusive$] meaning ``Audit inconclusive.''
        \end{description}
            \vspace{-.1in}
\end{enumerate}
            \vspace{-.1in}
\end{framed}
    \vspace{-.15in}
\caption{The competitive auditing game $\CompA$ with auditor
  $\CAuditor$, environment $\Environment$, election $E$, and advocate
  CVRs $\{ \cvr_\cA^{\sure} \mid \cA \in \candidates\}$.}
      \vspace{-.15in}
\label{fig:competitive-auditing-game}
\end{figure}

\begin{definition} [Disagreement Set] 
    Define the Omission Set as 
    $ \omission(\cvr_{\cA}, \cvr_{\cB}) = \Identifiers(\cvr_{\cA}) \setminus \Identifiers(\cvr_{\cB})$.

    Define the Conflict Set as $\conflict(\cvr_\cA, \cvr_\cB) = \{
        \iota \mid
        \text{$\iota \in \Identifiers(\cvr_{\cA}) \cap \Identifiers(\cvr_{\cB})$ and
          $\spred_{\cvr_{\cA}, \iota} \cap \spred_{\cvr_{\cB}, \iota} = \emptyset$}
        \}$
        
    The disagreement set is the set
      \[ \disagree(\cvr_\cA, \cvr_\cB) =  \substack{\omission(\cvr_\cA, \cvr_\cB)  \cup\\  \conflict(\cvr_\cA, \cvr_\cB)} \]
      The identifiers $\iota$ in the disagreement set are those for
      which $\cvr_\cA$ makes a claim about a ballot in $\mathbf{B}$ that
      is inconsistent with $\cvr_\cB$, in the sense that either 
      \begin{enumerate}
      \item 
      $\cvr_\cB$ does not recognize that the ballot exists at all $(\omission(\cvr_i, \cvr_j))$
      or 
      \item  $\cvr_\cA$ and $\cvr_\cB$ make
      contradictory claims about ballot interpretation $(\conflict(\cvr_\cA, \cvr_\cB))$.
      \end{enumerate}
\end{definition}
Intuitively, a CVR $\cvr_\cB$ will be disqualified if a ballot
$\ballot$ is found during the judgment phase with either i) an
identifier $\iota$ that was not reported on $\cvr_\cB$ or ii) an
interpretation $I^*$ that was not in the set of possible
interpretations $\spred_{\cvr_{\cA}, \iota}$.

\begin{definition} (Disqualification Function)
    Define function $\disqual: \Sigma^* \rightarrow \{0,1\}$ that takes in an identifier, $\iota$, an interpretation, $I^*$, and a CVR, $\cvr$. \[\disqual(\iota, I^*, \cvr) = \begin{cases} 1& {\scriptstyle \iota \not\in \Identifiers(\cvr) \vee I^* \not\in \spred_{\cvr, \iota}}\,,\\
    0 & \text{otherwise.}\end{cases}
\]    
\end{definition}
The intuition is that $\disqual$ looks up the identifier $\iota$ in $\cvr$ and outputs a disqualification vote if the provided interpretation $I^*$ is not present on the row $\cvr(\iota)$.

\noindent
We now analyze the competitive auditor $\CAuditor^t$ in
Figure~\ref{fig:competitive-auditor}. The parameter $t$ is a
  positive integer that calibrates the robustness of the audit to
  possible interpretation errors on the part of the advocates; see
  Theorems~\ref{thm:competitive-completeness-imperfect} and
  \ref{thm:competitive-soundness-imperfect}.

\begin{figure}[t!]
\small
  \begin{framed}
    \underline{Competitive auditor
      $\CAuditor^t[\candidates, \size; \cvr_1, \ldots,\cvr_k]$};
\medskip

The $\CAuditor^t$ immediately outputs $\inconclusive$ if any advocate announces duplicate labels.
    \medskip
    For each submitted CVR $\cvr_\cA$, define
$
      \Identifiers_\cA := \Identifiers(\cvr_\cA) = 
      \{ \iota \mid \text{the identifier $\iota$ appears in $\cvr_\cA$}\}
  $
    and, for an identifier $\iota \in \Identifiers_\cA$, let
    $\cinterpretation_{\cA,\iota}$ denote the set of interpretations associated
    with $\iota$ in $\cvr_\cA$.

    \begin{enumerate}[noitemsep]
    \item Remove from consideration any CVRs that do not have size
      $\size$.
    \item For each ordered pair $(\cA, \cB)$ for which the (remaining)
      CVRs $\cvr_\cA$ and $\cvr_\cB$ are contradictory:
    \begin{itemize}
        \item  Let $\iota^{(1)}, \ldots, \iota^{(t)}$ be independent and
      uniformly random elements of $\disagree(\cvr_\cA^{\sure}, \cvr_\cB^{\sure})$. For each
      $s \in \{1, \ldots, t\}$, request the ballot matching
      $\iota^{(s)}$. Let $I^{(1)},..., I^{(t)}$ denote the sequence of returned identifiers (including $\perp$).
      \item For each request $s$, we define the sequence $\disqual_s = \disqual(\iota^{(s)}, I^{(s)}, \cvr_\cB^{\sure})$.
      \item Finally, if $\sum_{s =1}^t \disqual_s > t/2$, disqualify $\cvr_\cB^{\sure}$.
    \end{itemize}
    \item  Remove all disqualified CVRs from consideration. 
    \item If there
    remains any pair of contradictory CVRs, or no remaining CVR
    declares victory, the audit concludes with $\inconclusive$.
    \item     Otherwise, a winner $\winner$ is declared by a remaining CVR that is not contradicted by any other remaining CVR and the audit concludes with ``$\winner$ is the winner.''
    \end{enumerate}
    \vspace{-.15in}
    \end{framed}
        \vspace{-.15in}
    \caption{The competitive auditor $\CAuditor^t$,
  with input $\candidates$, $\size$, and advocate
  CVRs $\{ \cvr_\cA^{\sure} \mid \cA \in \candidates\}$ where there are $k$ submitted CVRs. The parameter $t$ determines the number of
  ballot samples collected by the auditor: in particular, the auditor
  requests no more than $t k (k-1)$ ballots.}
              \vspace{-.1in}
  \label{fig:competitive-auditor}
\end{figure}

\subsection{Analysis of the Competitive Auditor $\CAuditor^t$}
In this section, we make two claims arguing about the completeness and soundness of $\CAuditor$.  We first consider these claims in the case that an ``honest'' party makes no errors and then consider the setting with errors in the next subsection.

\textbf{Theorem~\ref{thm:competitive-completeness-simple}} shows that if
there is an election winner $\cW$ and some $\cvr_{\cW}$ is consistent
with the physical ballots, then $\cW$ will be the output of
$\CAuditor$.  This theorem assumes an honest environment. Recall that
(i.) the consistency of a CVR means that for each ballot, the CVR
submits a superset of the actual interpretations for that ballot and
(ii.) a candidate is an election winner only if it wins regardless of
the interpretation of ballots with multiple interpretations.

\textbf{Theorem~\ref{thm:competitive-soundness-simple}} shows that if a
candidate is an election loser $\loser$ and one of the submitted $\cvr$s is 
canonical,
then $\loser$ will never be the output of $\CAuditor$.  This theorem
does not require an honest environment. Recall that a CVR is canonical
if for every ballot it declares exactly the set of interpretations on
the physical ballots.  Intuitively, since a loser loses no matter the
interpretation declared by the environment, a canonical CVR is never
disqualified by $\CAuditor$ and $\loser$ can never be the output.

The main requirement for both theorems is that some advocate asserts
the full set of interpretations for each physical ballot on the CVR;
this may be either a superset or the exact set, depending on the
circumstances. An advocate is disqualified when a ballot is delivered
with a ground-truth interpretation that does not appear in the
advocate's CVR. Note that any ballot for which no ground-truth
interpretation appears on the CVR is sure to disqualify the advocate
and, in fact, the analysis only requires this weaker statement.

\begin{theorem}[Competitive completeness]
  \label{thm:competitive-completeness-simple} Let $\CAuditor^t$ be the
  competitive auditor with sample parameter $t\in\mathbb{Z}^+$. Let
  $E = (\candidates, \mathbf{B}, \size)$ be an election with winner
  $\winner \in \candidates$. Let $\cvr_*$ be consistent with $E$ and
  declare $\winner$ to be the winner. Let
  $\cvr_1, \ldots, \cvr_{k-1}$ be any collection of CVRs. Then
  \[
    \CompA[\CAuditor^t,
    \Environment; E; \cvr_*, \cvr_1, \ldots, \cvr_{k-1}] = \winner\,,
  \]
  for any honest environment $\Environment$. Furthermore, $\CAuditor^t$
  completes the audit with no more than $t k (k-1)$ ballot samples.
\end{theorem}

We remark that the conclusion of the Theorem holds even for $t=1$, in
which case that auditor only draws $k(k-1)$ ballots. Larger values of $t$
are relevant for managing interpretation errors; see the next
subsection.

\ifnum\addproofs=1
\begin{proof}[Proof of Theorem~\ref{thm:competitive-completeness-simple}]
  Adopt notation as described in the statement of the theorem: an
  election $E = (\candidates, \size, \mathbf{B})$ with uniquely
  labeled ballots and winner $\winner$, and a honest environment
  $\Environment$. Let $\cvr_*$ be a CVR consistent with $E$, also
  declaring winner $\winner$, and let $\cvr_1, \ldots, \cvr_{k-1}$ be
  any sequence of CVRs. Considering that $\cvr_*$ is consistent with
  $E$, it is not possible for $\cvr_*$ to amass any votes for
  disqualification during comparison with any alternative CVR, as by
  definition it provides a consistent interpretation for any ballot
  $\mathbf{b}$ and interpretation $I \in \struth_\mathbf{b}$
  sampled by $\CAuditor$. It follows that the only possible outcomes
  of the audit are \inconclusive\ or ``$\winner$ is the winner.''

  To complete the proof, we establish that any CVR $\cvr_i$ that
  declares $\winner$ to be a loser---which is to say that there is a
  candidate $\cA$ for which
  $\cvr_i^{\vfloor}(\cA) > \cvr_i^{\vceil}(\cW)$---will necessarily be
  disqualified (by $\cvr_*$) during the audit; this ensures that no
  contradictory pairs of CVRs remain at the conclusion of the audit.

  We establish that the set $\Claim_{i\prec *}$ (associated with the
  two CVRs $\cvr_i$ and $\cvr_*$) cannot be empty: Observe that if
  $\Claim_{i\prec *}$ were, in fact, empty then the two CVRs would
  declare the same set of identifiers and, moreover, there would be an
  interpretation $I_{\mathbf{b}}^\sim$ for each ballot $\mathbf{b}$
  consistent with both CVRs in the sense that
  $I_{\mathbf{b}}^\sim \in \spred_{\cvr_{*}, \iota} \cap
  \spred_{\cvr_{i}, \iota}$. But, considering the contradictory claims
  made by the two CVRs, this would imply that both
  \[
    \sum_{\mathbf{b}} I^\sim_{\iota_\mathbf{b}}(\cA) \leq    \cvr_*^{\vceil}(\cA)     < \cvr_*^{\vfloor}(\winner) \leq \sum_{\mathbf{b}} I^\sim_{\iota_\mathbf{b}}(\winner)
  \]
  and
  \[
    \sum_{\mathbf{b}} I^\sim_{\iota_\mathbf{b}}(\winner) \leq    \cvr_i^{\vceil}(\winner)     < \cvr_i^{\vfloor}(\cA) \leq \sum_{\mathbf{b}} I^\sim_{\iota_\mathbf{b}}(\cA)\,,
  \]
  a contradiction. It follows that $\Claim_{i\prec *}$ is nonempty. As
  we assume that $\Environment$ is honest, any identifier requested
  during the audit from $\Claim_{i\prec *}$ will be properly serviced
  by the environment and will lead to a vote for disqualification for
  $\cvr_i$. In this case, with a consistent $\cvr_*$ and honest
  $\Environment$, all $t$ votes collected during consideration of the
  pair $(i,*)$ will call for disqualification. Thus $\cvr_i$ will be
  removed from consideration. As all contradictory CVRs will be
  so removed, the audit will conclude with ``$\winner$ is the winner,''
  as desired.
\end{proof}
\fi

\begin{theorem}[Competitive soundness]
  \label{thm:competitive-soundness-simple}
  Let $\CAuditor^t$ be the
  competitive auditor with integer sample parameter $t > 0$. Let
  $E = (\candidates, \mathbf{B}, \size)$ be an election with loser
  $\loser \in \candidates$. Let $\cvr_*$ be the canonical CVR for
  $E$. Let $\cvr_1, \ldots, \cvr_{k-1}$ be any collection of CVRs. Then
  \[
    \Pr\bigl[\CompA[\CAuditor, \Environment; E; \cvr_*, \cvr_1, \ldots,
    \cvr_{k-1}\bigr] = \loser\bigr] = 0\,,
  \]
  for any environment $\Environment$. Furthermore, $\CAuditor^t$
  completes the audit with no more than $t k (k-1)$ ballot samples.
\end{theorem}

\ifnum\addproofs=1
\begin{proof}[Proof of Theorem~\ref{thm:competitive-soundness-simple}]
  Adopt notation as described in the statement of the theorem: an
  election $E = (\candidates, \size, \mathbf{B})$ with uniquely
  labeled ballots and losing candidate $\loser$.  We now consider an
  arbitrary environment $\Environment$. Let $\cvr_*$ be the canonical
  CVR and let $\cvr_1, \ldots, \cvr_{k-1}$ be any sequence of
  CVRs. Considering that $\cvr_*$ is canonical with $E$, it is not
  possible for $\cvr_*$ to amass any votes for disqualification during
  comparison with any alternative CVR, as by definition it provides a
  consistent interpretation for any ballot $\mathbf{b}$ and
  interpretation $I^* \in \struth_\mathbf{b}$ sampled by
  $\CAuditor$. Note, additionally, that as $\cvr_*$ is the canonical
  CVR, it reflects the fact that $\loser$ is a loser: there is a
  candidate $\cA$ so that
  $\cvr_*^{\vceil}(\loser) < \cvr_*^{\vfloor}(\cA)$. In particular,
  $\cvr_*$ contradicts any CVR declaring $\loser$ to be the winner. As
  $\cvr_*$ cannot be disqualified, the audit cannot conclude with
  ``$\loser$ is the winner,'' as desired.
\end{proof}
\fi

\subsection{Handling Advocacy Errors}
\label{sec:contested with errors}

As producing cast-vote records is a process involving manipulation and
interpretation of physical ballots, it may be unrealistic to demand
that even a well-intentioned auditor can produce a perfectly
consistent (or canonical) CVR.
Occasional mistakes may occur during ballot interpretation, there may
be some uncertainty about establishing a set $\mathtt{I}$ that
necessarily contains all interpretations that an auditing board may
assign to a ballot, and ballots may be overlooked. As we insist in our
setting that CVRs have size consistent with the election, we note that
an advocate can always pad out a CVR by adding invented identifiers
that do not correspond to any ballot. All told, these considerations
lead us to consider CVRs that are ``nearly'' consistent, defined
formally below.

Theorems~\ref{thm:competitive-completeness-imperfect} and \ref{thm:competitive-soundness-imperfect} extend these theorems to the case when ``honest'' parties make a small fraction of errors.

\begin{definition} Let $E = (\candidates, \size, \mathbf{B})$ be an
  election.  We say that a $\cvr$ (over $\candidates$) is
  $(1 - \epsilon)$-consistent if the size of the CVR is $\size$ and
  there is a subset $C$ of $\Identifiers(\cvr)$ of size at least
  $(1 - \epsilon) \size$ on which $\cvr$ is consistent, which is to
  say that for each $\iota \in C$, there is a ballot
  $\mathbf{b} \in \mathbf{B}$ for which $\iota = \iota_{\mathbf{b}}$
    and $\mathtt{I}_{\mathbf{b}} \subset \mathtt{I}_\iota$ (where
    $\mathtt{I}_\iota$ is the set of interpretations declared by
    $\cvr$ for $\iota$). Define a $(1-\epsilon)$-canonical $\cvr$ analogously.
\end{definition}

 Let $\Binomial[\gamma,t;t/2]$ denotes the tail of the binomial
  distribution: specifically, if $X_1, \ldots, X_t$ are independent
  Bernoulli random variables for which $\Pr[X_i = 1] = \gamma$, then
  $\Binomial[\gamma,t;t/2]$ is the probability that
  $\sum_i X_i \geq t/2$.

\begin{theorem}[Competitive completeness with imperfect advocacy]
  \label{thm:competitive-completeness-imperfect}
  Let $\CAuditor^t$ be the competitive auditor with integer sample
  parameter $t > 0$. Let $E = (\candidates, \mathbf{B}, \size)$ be an
  election with winner $\winner \in \candidates$. Let $\cvr_*$ be
  $(1-\epsilon)$-consistent with $E$ and declare $\winner$ to be the
  winner with margin $\mu_{\cvr_*} > 4 \epsilon$. Let
  $\cvr_1, \ldots, \cvr_{k-1}$ be any collection of CVRs and let
  $\Environment$ be an honest environment. Then
  \begin{align*}
    \Pr&[\CompA[\CAuditor^t, \Environment; E; \cvr_*, \cvr_1, \ldots,
    \cvr_{k-1}] = \winner] \\&\geq 1 - 2(k-1) \Binomial[2\epsilon/\mu_{\cvr_*},t;t/2]\,.
  \end{align*}
\end{theorem}

\noindent
For $\gamma < 1/2$ (guaranteed above by
$4 \epsilon < \mu_{\cvr_*}$), the quantity
$\Binomial[\gamma,t;t/2] = \exp(-\omega(t))$. Under the mild assumption that $\epsilon < \mu_{\cvr_*} / 6$,
say, the number of samples required by such an audit to achieve any
particular risk limit is independent of $\mu_{\cvr_*}$.  An advocate may reduce $\mu_{\cvr_*}$ in an effort to be consistent with the underlying ballots.

\ifnum\addproofs=1
\begin{proof}[Proof of Theorem~\ref{thm:competitive-completeness-imperfect}]
  Adopt notation as described in the statement of the theorem: an
  election $E = (\candidates, \size, \mathbf{B})$ with uniquely
  labeled ballots and winner $\winner$, and a honest environment
  $\Environment$. Let $\cvr_*$ be $(1-\epsilon)$-consistent with $E$,
  declaring winner $\winner$ with margin $\mu_* > 4 \epsilon$; let
  $\cvr_1, \ldots, \cvr_{k-1}$ be any sequence of CVRs. Consider a
  CVR $\cvr_i$ that contradicts $\cvr_*$; in this case
  $\cvr_i^{\vfloor}(\cA) > \cvr_i^{\vceil}(\winner)$ for a candidate $\cA$. 
  By assumption, 
  $\cvr_*^{\vfloor}(\winner) \geq \cvr_*^{\vceil}(\cA) + \size \cdot \mu_*$.  It follows that both $\Claim_{i\prec *}$ and $\Claim_{*\prec i}$
  must have cardinality at least $\size \cdot \mu_*/2$.

  There are two events of interest: the event that $\cvr_*$ is
  disqualified by some contradictory $\cvr_i$, and the event that some
  contradictory $\cvr_i$ is not disqualified by $\cvr_*$. 
  
  For the
  first of these, observe that while
  $|\Claim_{* \prec i}| \geq \size \mu_* /2$ for a contradictory
  $\cvr_i$, the CVR $\cvr_*$ is consistent with the actual ballots on
  all but $\epsilon \size$ identifiers. It follows that the
  probability that an identifier chosen uniformly from
  $\Claim_{* \prec i}$ will reference a ballot that is inconsistent
  with $\cvr_*$ is no more than $\epsilon / (\mu_*/2) = 2\epsilon/\mu_*$. Thus the
  probability that $\cvr_*$ is disqualified by $\cvr_i$ is no more
  than $\Binomial[2\epsilon/\mu_*,t;t/2]$, as desired. As there are no
  more than $k-1$ contradictory CVRs, the probability that $\cvr_*$ is
  disqualified is no more than
  $(k-1)\Binomial[2\epsilon/\mu_*,t;t/2]$. 
  
  The second event can be
  bounded by similar means. Consider a contradictory $\cvr_i$;
  as above, the set $\Claim_{i\prec *}$ has size at least
  $\size\cdot\mu_*/2$ and, of these identifiers, all but $\epsilon \size$
  of them are consistently reflected by the ballots. It follows that
  the probability that a random identifier drawn from
  $\Claim_{i\prec *}$ will be consistent with the underlying ballots
  (and hence generate a vote for disqualification for $\cvr_i$
  considering that $\Environment$ is honest) is at least
  $1-\epsilon/(\mu/2)$. Recall that the the binomial distribution
  corresponding to the sum of $t$ i.i.d.\ Bernoulli random variables
  with expectation $1-\gamma$ is precisely the distribution
  corresponding to $t$ variables with expectation $\gamma$ reflected
  about the point $t/2$.  Thus such a contradictory CVR will avoid
  disqualification with probability no more than
  $\Binomial[2\epsilon/\mu_*,t;t/2]$. As there are no more than $k-1$
  contradictory CVRs, the probability that any of them are not
  disqualified is no more than
  $(k-1) \Binomial[2\epsilon/\mu_*,t;t/2]$.

  Combining these two bounds yields the final estimate, as in this
  case any contradictory CVR will have been disqualified while
  $\cvr_*$ will remain.
\end{proof}
\fi

\begin{theorem}[Competitive soundness] 
\label{thm:competitive-soundness-imperfect}
Let $\CAuditor^t$ be the
  competitive auditor with integer sample parameter $t > 0$. Let
  $E = (\candidates, \mathbf{B}, \size)$ be an election with loser
  $\loser \in \candidates$. Let $\cvr_*$ be a $(1-\epsilon)$-canonical CVR for $E$. 
  Define \[\mu_{\cvr_*} = \frac{\max_{\cA \in \candidates \setminus \loser}   (\cvr_*^-(\cA) -\cvr_*^+(\loser))}{\size}.\]
 Let
  $\cvr_1, \ldots, \cvr_{k-1}$ be any collection of CVRs. Then
  \begin{align*}
    \Pr&\bigl[\CompA[\CAuditor^t, \Environment; E; \cvr_*, \cvr_1,
    \ldots, \cvr_{k-1}\bigr] = \loser\bigr] \\&\leq
    2(k-1)\Binomial[2\epsilon/\mu,t;t/2]
  \end{align*}
  for any environment $\Environment$. Furthermore, $\CAuditor^t$
  completes the audit with no more than $t k (k-1)$ ballot samples.
\end{theorem}

\ifnum\addproofs=1
\begin{proof}
  Follows from the same reasoning as Theorem~\ref{thm:competitive-completeness-imperfect}.
\end{proof}
\fi

\subsection{Remarks on Bayesian competitive audits}
\label{sec:bayesian contested}
While we do not present any details, we make a few remarks about
competitive audits in the Bayesian case. This calls for each advocate to
submit a Bayesian CVR. Now we note that identification of a collection
of ballots $D$ on which the two CVRs substantively disagree permits us
to define two different probability distributions on this set of
ballots: (i.) select a ballot at random from $D$, and (ii.) output an
interpretation given by the CVR in question. Then the classical
sequential probability ratio test provides a statistical test for
distinguishing the two models~\cite{wald1992sequential}. This can
serve as the statistical mechanism for selecting between contradictory
CVRs. \ifnum\conference=0 One can achieve a constant number of ballots
  to disqualify a CVR only if there are a constant number of ballots
  whose respective $D$ have constant Kullback-Leibler-divergence, this
  is why we treat the simpler case where a constant number of ballots
  have disjoint interpretations.  \fi

\section*{Acknowledgements}
These results were developed as part of a collaboration with the
Office of the Connecticut Secretary of State and, additionally, were supported
in part by a grant from that office.  Discussions with anonymous
reviewers improved the narrative and technical treatment. The authors also thank
Ron Rivest for his helpful comments.  B.F. is
supported by NSF Grants \#2232813 and \#2141033 and the Office of
Naval Research. A.R was supported by NSF Grant \#1801487.



\small
\bibliographystyle{unsrt}
\bibliography{../RLA}

\begin{thebibliography}{10}

\bibitem{lindeman2012gentle}
Mark Lindeman and Philip~B Stark.
\newblock A gentle introduction to risk-limiting audits.
\newblock {\em IEEE Security \& Privacy}, 10(5):42--49, 2012.

\bibitem{national2018securing}
National~Academies of~Sciences~Engineering and Medicine.
\newblock {\em Securing the Vote: Protecting American Democracy}.
\newblock National Academies Press, 2018.

\bibitem{zagorski2021minerva}
Filip Zag{\'o}rski, Grant McClearn, Sarah Morin, Neal McBurnett, and Poorvi~L
  Vora.
\newblock Minerva--an efficient risk-limiting ballot polling audit.
\newblock In {\em {USENIX} Security Symposium}, pages 3059--3076. {USENIX}
  Association, 2021.

\bibitem{bajcsy2015systematic}
Andrea Bajcsy, Ya-Shian Li-Baboud, and Mary Brady.
\newblock Systematic measurement of marginal mark types on voting ballots,
  2015.
\newblock \url{https://nvlpubs.nist.gov/nistpubs/ir/2015/NIST.IR.8069.pdf}.

\bibitem{harrison2022adaptive}
Benjamin Fuller, Abigail Harrison, and Alexander Russell.
\newblock Adaptive risk-limiting comparison audits.
\newblock In {\em IEEE Symposium on Security and Privacy}, pages 2002--2019,
  Los Alamitos, CA, USA, may 2023.

\bibitem{stark2009auditing}
Philip~B Stark.
\newblock Auditing a collection of races simultaneously.
\newblock {\em arXiv preprint arXiv:0905.1422}, 2009.

\bibitem{stark2010super}
Philip~B Stark.
\newblock Super-simple simultaneous single-ballot risk-limiting audits.
\newblock In {\em EVT/WOTE}, 2010.

\bibitem{higgins2011sharper}
Michael~J Higgins, Ronald~L Rivest, and Philip~B Stark.
\newblock Sharper p--values for stratified election audits.
\newblock {\em Statistics, Politics, and Policy}, 2(1), 2011.

\bibitem{ottoboni2018risk}
Kellie Ottoboni, Philip~B Stark, Mark Lindeman, and Neal McBurnett.
\newblock Risk-limiting audits by stratified union-intersection tests of
  elections ({SUITE}).
\newblock In {\em International Joint Conference on Electronic Voting}, pages
  174--188. Springer, 2018.

\bibitem{stark2020sets}
Philip~B Stark.
\newblock Sets of half-average nulls generate risk-limiting audits: Shangrla.
\newblock In {\em Financial Cryptography and Data Security}, pages 319--336.
  Springer, 2020.

\bibitem{waudby2021rilacs}
Ian Waudby-Smith, Philip~B Stark, and Aaditya Ramdas.
\newblock Rilacs: Risk limiting audits via confidence sequences.
\newblock In {\em International Joint Conference on Electronic Voting}, pages
  124--139. Springer, 2021.

\bibitem{blom2021assertion}
Michelle Blom, Jurlind Budurushi, Ronald~L Rivest, Philip~B Stark, Peter~J
  Stuckey, Vanessa Teague, and Damjan Vukcevic.
\newblock Assertion-based approaches to auditing complex elections, with
  application to party-list proportional elections.
\newblock In {\em International Joint Conference on Electronic Voting}, pages
  47--62. Springer, 2021.

\bibitem{CO-DiscReport}
Colorado~Secretary of~State.
\newblock 2020 general election risk-limiting audit discrepancy report, 2020.
\newblock
  \url{https://www.sos.state.co.us/pubs/elections/RLA/2020/general/DiscrepancyReport.pdf}.

\bibitem{VoTeR2022}
A.~Russell, L.~Michel, B.~Fuller, J.~Wohl, and G.~Johnson.
\newblock Statistical analysis of post-election audit data for the {November}
  8, 2022 state election.
\newblock
  \url{https://voter.engr.uconn.edu/wp-content/uploads/sites/3651/2023/02/2022-11-08-hand-count-statistics.pdf}.

\bibitem{wang2010openscan}
Kai Wang, Eric Rescorla, Hovav Shacham, and Serge~J Belongie.
\newblock Openscan: A fully transparent optical scan voting system.
\newblock {\em EVT/WOTE}, 10:1--13, 2010.

\bibitem{starkconservative}
Philip~B. Stark.
\newblock {Conservative statistical post-election audits}.
\newblock {\em The Annals of Applied Statistics}, 2(2):550 -- 581, 2008.

\bibitem{bernhard2021risk}
Matthew Bernhard.
\newblock Risk-limiting audits: A practical systematization of knowledge.
\newblock In {\em International Joint Conference on Electronic Voting}, 2021.

\bibitem{verifiedvotingprinciples}
Lynn Garland, Neal McBurnett, Jennier Morrell, Marian~K. Schneider, and
  Stephanie Singer.
\newblock Principles and best practices for post-election tabulation audits,
  2018.

\bibitem{Hall2009}
Joseph~Lorenzo Hall, Luke~W Miratrix, Philip~B Stark, Melvin Briones, Elaine~\
  Ginnold, Freddie Oakley, Martin Peaden, Gail Pellerin, Tom Stanionis, and
  Tricia Webber.
\newblock {Implementing risk-limiting post-election audits in California.}
\newblock In {\em {EVN/WOTE}}, Montreal, Canada, 2009.

\bibitem{lindeman2012bravo}
Mark Lindeman, Philip~B Stark, and Vincent~S Yates.
\newblock Bravo: Ballot-polling risk-limiting audits to verify outcomes.
\newblock In {\em EVT/WOTE}, 2012.

\bibitem{checkoway}
Stephen Checkoway, Anand Sarwate, and Hovav Shacham.
\newblock Single-ballot risk-limiting audits using convex optimization.
\newblock In {\em Proceedings of the 2010 International Conference on
  Electronic Voting Technology/Workshop on Trustworthy El\ ections},
  EVT/WOTE'10, pages 1--13, USA, 2010. USENIX Association.

\bibitem{starkcast}
Philip~B. Stark.
\newblock Cast: Canvass audits by sampling and testing.
\newblock {\em IEEE Transactions on Information Forensics and Security},
  4(4):708--717, 2009.

\bibitem{blom2022first}
Michelle Blom, Peter~J Stuckey, Vanessa Teague, and Damjan Vukcevic.
\newblock A first approach to risk-limiting audits for single transferable vote
  elections.
\newblock In {\em International Conference on Financial Cryptography and Data
  Security}, pages 366--380. Springer, 2022.

\bibitem{stark2009efficient}
Philip~B Stark.
\newblock Efficient post-election audits of multiple contests: 2009 california
  tests.
\newblock In {\em CELS 2009 4Th annual conference on empirical legal studies
  paper}, 2009.

\bibitem{ottoboni2019bernoulli}
Kellie Ottoboni, Matthew Bernhard, J~Alex Halderman, Ronald~L Rivest, and
  Philip~B Stark.
\newblock Bernoulli ballot polling: a manifest improvement for risk-limiting
  audits.
\newblock In {\em International Conference on Financial Cryptography and Data
  Security}, pages 226--241, 2019.

\bibitem{Stark:Conservative}
Philip~B. Stark.
\newblock Risk-limiting postelection audits: Conservative {$P$}-values from
  common probability inequalities.
\newblock {\em IEEE Transactions on Information Forensics and Security},
  4(4):1005--1014, 2009.

\bibitem{stark2023alpha}
Philip~B Stark.
\newblock Alpha: audit that learns from previously hand-audited ballots.
\newblock {\em The Annals of Applied Statistics}, 17(1):641--679, 2023.

\bibitem{banuelos2012limiting}
Jorge~H Banuelos and Philip~B Stark.
\newblock Limiting risk by turning manifest phantoms into evil zombies.
\newblock {\em arXiv preprint arXiv:1207.3413}, 2012.

\bibitem{zagorski2020athena}
Filip Zag{\'o}rski, Grant McClearn, Sarah Morin, Neal McBurnett, and Poorvi~L
  Vora.
\newblock The athena class of risk-limiting ballot polling audits.
\newblock {\em arXiv preprint arXiv:2008.02315}, 2020.

\bibitem{RLAWorkbook}
Jennifer Morrell.
\newblock Knowing it's right, part two. risk-limiting audit implementation
  workbook., 2019.

\bibitem{rla-working-group}
Jennie Bretschneider, Sean Flaherty, Susannah Goodman, Mark Halvorson, Roger
  Johnston, Mark Lindeman, Ronald~L. Rivest, Pam Smith, and Phillip~B. Stark.
\newblock Risk-limiting post-election audits: Why and how, 2012.

\bibitem{glazer2020bayesian}
Amanda~K Glazer, Jacob~V Spertus, and Philip~B Stark.
\newblock Bayesian audits are average but risk-limiting audits are above
  average.
\newblock In {\em E-Vote-ID 2020}, pages 84--94. Springer, 2020.

\bibitem{benaloh2021vault}
Josh Benaloh, Kammi Foote, Philip~B Stark, Vanessa Teague, and Dan~S Wallach.
\newblock {VAULT}-style risk-limiting audits and the {I}nyo county pilot.
\newblock {\em IEEE Security \& Privacy}, 19(4):8--18, 2021.

\bibitem{jones2022scan}
Douglas~W Jones, Sunoo Park, Ronald~L Rivest, and Adam Sealfon.
\newblock Scan, shuffle, rescan: Machine-assisted election audits with
  untrusted scanners.
\newblock In {\em Financial Cryptography}, 2024.

\bibitem{reif1984complexity}
John~H Reif.
\newblock The complexity of two-player games of incomplete information.
\newblock {\em Journal of computer and system sciences}, 29(2):274--301, 1984.

\bibitem{Kiwi:2000aa}
M.~Kiwi, C.~Lund, D.~Spielman, A.~Russell, and R.~Sundaram.
\newblock Alternation in interaction.
\newblock {\em Computational Complexity}, 9(3):202--246, 2000.

\bibitem{babai1991non}
L{\'a}szl{\'o} Babai, Lance Fortnow, and Carsten Lund.
\newblock Non-deterministic exponential time has two-prover interactive
  protocols.
\newblock {\em Computational complexity}, 1:3--40, 1991.

\bibitem{pai2024questionable}
Rashmi Pai.
\newblock Marginal mark {RLA} tools.
\newblock \url{https://github.com/rpai0005/Questionable-Simulation-Tools/},
  2024.

\bibitem{harrison2023adaptive}
Abigail Harrison.
\newblock Adaptive {RLA} tools.
\newblock \url{https://github.com/aeharrison815/Adaptive-RLA-Tools}, 2023.

\bibitem{spertus2023cobra}
Jacob~V Spertus.
\newblock Cobra: Comparison-optimal betting for risk-limiting audits.
\newblock In {\em International Conference on Financial Cryptography and Data
  Security}, pages 95--109. Springer, 2023.

\bibitem{wald1992sequential}
Abraham Wald.
\newblock Sequential tests of statistical hypotheses.
\newblock In {\em Breakthroughs in statistics: Foundations and basic theory},
  pages 256--298. Springer, 1992.

\end{thebibliography}

\appendix

\end{document}